\documentclass{aa}

\usepackage{graphicx}
\usepackage{txfonts}
\usepackage{multirow}
\newcommand{\mr}{\multirow{2}{*}}
\newcommand{\mc}{\multicolumn{2}{c}}
\newcommand{\fmp}{\texttt{FMP}}
\usepackage[breaklinks=true,colorlinks=true,citecolor=blue,linkcolor=blue,urlcolor=blue]{hyperref}
\usepackage{tablefootnote}

\begin{document}

\title{MUSE adaptive-optics spectroscopy confirms dual active galactic nuclei and strongly lensed systems at sub-arcsec separation}


\titlerunning{Dual AGN classification with MUSE}
\authorrunning{Scialpi et al.}

\author{M. Scialpi \inst{1,2,3}
\and F. Mannucci \inst{3}
\and C. Marconcini \inst{2,3}
\and G. Venturi \inst{5,3}
\and E. Pancino \inst{3}
\and A. Marconi \inst{2,3}
\and G. Cresci \inst{3}
\and F. Belfiore \inst{3}
\and A. Amiri \inst{2,6}
\and E. Bertola \inst{3}
\and S. Carniani \inst{5}
\and C. Cicone \inst{7} 
\and A. Ciurlo\inst{8}
\and Q. D'Amato \inst{3}
\and M. Ginolfi \inst{2,3,9}
\and E. Lusso \inst{2,3}
\and A. Marasco \inst{10}
\and E. Nardini \inst{3}
\and K. Rubinur\inst{7}
\and P. Severgnini \inst{11}
\and G. Tozzi \inst{2,3}
\and L. Ulivi \inst{1,2,3}
\and C. Vignali \inst{12}
\and M. Volonteri \inst{13}
}
  
\institute{
University of Trento, Via Sommarive 14, I-38123 Trento, Italy \\ 
\email{martina.scialpi@unitn.it}
\and 
Università di Firenze, Dipartimento di Fisica e Astronomia, via G. Sansone 1, 50019 Sesto F.no, Firenze, Italy \\ \email{martina.scialpi@unifi.it}
\and 
INAF - Osservatorio Astrofisico di Arcetri, Via Largo E. Fermi 5, 50125 Firenze, Italy \\ 
\email{martina.scialpi@inaf.it}
\and  
Instituto de Astrofísica, Facultad de Física, Pontificia Universidad Católica de Chile, Casilla 306, Santiago 22, Chile
\and 
Scuola Normale Superiore, Piazza dei Cavalieri 7, 56126, Pisa, Italy.
\and 
Department of Physics, University of Arkansas, 226 Physics Building, 825 West Dickson Street, Fayetteville, AR 72701, USA
\and 
Institute of Theoretical Astrophysics, University of Oslo, P.O Box 1029, Blindern, 0315 Oslo, Norway
\and 
Department of Physics and Astronomy, University of California Los Angeles, 430 Portola Plaza, Los Angeles, CA 90095, USA 
\and 
European Southern Observatory, Karl-Schwarzschild-Str. 2, D-85748 Garching, Germany
\and 
INAF - Padova Astronomical Observatory, Vicolo dell’Osservatorio 5, 35122 Padova, Italy
\and 
INAF - Osservatorio Astronomico di Brera, via Brera 28, 20121, Milano, Italy
\and 
Physics and Astronomy Department "Augusto Righi", Università di Bologna, Via Gobetti 93/2, 40129 Bologna, Italy.
\and 
Institut d'Astrophysique de Paris,  98bis Bd Arago, 75014 Paris, France.
}
   \date{}

  \abstract
   {
   The novel Gaia multi peak (GMP) technique has proven to be able to successfully select dual and lensed active galactic nuclei (AGN) candidates at sub-arcsecond separations. Both populations are important because dual AGN represent one of the central, still largely untested, predictions of $\Lambda$CDM cosmology, and compact lensed AGN allow us to probe the central regions of the lensing galaxies.
   In this work, we present high-spatial-resolution spectroscopy of 12 GMP-selected systems. We used the adaptive-optics assisted integral-field spectrograph
   MUSE at the VLT
  to resolve each system and investigate the nature of each component.  
   All targets show the presence of two components confirming the GMP selection. We classify 4 targets as dual AGN, 3 as lensed quasar candidates, and 5 as a chance alignment of a star and an AGN. With separations ranging from 0.30$''$ to 0.86$''$, these dual and lensed systems are among the most compact systems discovered to date at $z >0.5$. This is the largest sample of distant dual AGN with sub-arcsecond separations ever presented in a single paper.
}

   \keywords{Galaxies: active --
quasars: Dual,  --
Gravitational lensing: general}

   \maketitle
%

 \section{Introduction}

Supermassive black hole (SMBH) pairs at sub-kiloparsec scales ($\sim0.5\arcsec$) are expected to form at the center of galaxies during the hierarchical assembly of structures \citep[e.g.,][]{Begelman80, Mayer07}. During galaxy mergers, SMBH pairs can undergo orbital decay via dynamical friction, form a gravitationally bound binary 
\citep[e.g.,][among many others]{Kelley17,Volonteri21},
and eventually coalesce with the associated emission of gravitational waves (GWs), as was predicted by Albert Einstein's general theory of relativity \citep{Einstein36}. Dual active galactic nuclei (AGN) are systems containing two active nuclei powered by accretion onto two different SMBHs that are nested inside their host, but that are not yet mutually gravitationally bound. These systems are the natural precursors of coalescing binary SMBHs, which are strong emitters of low-frequency GWs \citep{Colpi14}.  
Quantifying their overall abundance would help us to estimate the GW event rate in the future ESA mission Laser Interferometer Space Antenna (LISA), which  will unveil the rich population of SMBHs of $\sim 10^{4-7}\ M_\odot$ forming in collisions out to redshifts as large as $z\sim 20$ \citep[e.g.,][]{Amaro-Seoane23, Arzoumanian18}.
Moreover, dual AGN are relevant to understanding the theories of hierarchical structure formation, growth, and demography of SMBHs, AGN fuelling and feedback, and the role of merging in AGN evolution \citep{Ellison14, Perna23, Sedda23}.
The discovery of two nearby AGN at the same redshift can reveal the presence of a dual system (two different SMBHs), but can also be due to the lensing of a single quasar (QSO) by an intervening foreground galaxy. 
Lensed systems at small projected separations are a powerful tool with which to address key questions in astrophysics, like studying intrinsically faint and distant lensed sources, investigating dark matter properties of the deflector \citep{Massey10}, and providing constraints on fundamental cosmological quantities, like the Hubble constant, the cosmological constant, and the density parameter of the Universe \citep{Wong20,Lemon23}. These studies are possible since many properties of the lensed systems depend on the age, scale, and overall geometry of the Universe.
Therefore, both dual and lensed AGN are very powerful tools with which to answer a wide range of open questions.

The direct observation of dual and lensed AGN at such small separations is challenging due to the combination of their rarity and the limited spatial resolutions  of current wide-field facilities.
The ESA Gaia satellite \citep{Prusti16} has been used in various ways to select these elusive systems. 
For example, dual and lensed systems have been discovered by looking for multiple Gaia sources associated, within a few arcsecond, with known AGN \citep{Lemon17, Chen22a, Shen23b}. 
In addition, source variability can also be exploited to find multiple AGN, as was proposed by \cite{Kochanek06} and employed by \cite{KroneMartins19}, among others.
More compact systems have been identified by looking for the astrometric jitter associated with AGN variability in unresolved dual systems \citep[“varstrometry”;][]{Hwang20, Shen19, Gross23}.
Despite all these efforts, until recently, very few dual AGN with separations below $\sim 7 $ kpc have been confirmed at $z>0.5$ (e.g., \citealt{Lemon19, Chen22b, Chen23a, Glikman23}, see \citealt{Mannucci23} for a full list), making any quantitative testing of cosmological predictions impossible, especially at high redshift.

Recently, a novel technique dubbed “Gaia multi-peak” (GMP; \citealt{Mannucci22}) was developed by our group to obtain large and reliable samples of dual and lensed AGN candidates with sub-arcsecond separations. This method looks for AGN that present multiple peaks in the $1D$ light profiles observed by Gaia.
The GMP method consists of selecting AGN with large values of the parameter \texttt{ipd\_frac\_multi\_peak} (\fmp) provided in the Gaia catalog since the Early Data Release 3 (EDR3, \citealt{Brown21}). This parameter  gives the fraction of differently oriented scans in which the object appears to have multiple light peaks. The statistical properties of the GMP selection have been described in \cite{Mannucci22} and \cite{Mannucci23}.

GMP-selected objects could be physically associated dual AGN, gravitationally lensed systems, or an AGN projected close to a foreground, Galactic star. Therefore, accurate, spatially resolved spectroscopy is necessary to unveil the nature of each system. Integral field spectroscopy is particularly well-suited to extract spatially resolved spectra of each component of these systems because, in most cases, the positions of the various components are not known a priori. Moreover, either space observations or the use of adaptive optics (AO) from the ground are needed due to the small angular separations of the components. The first spatially resolved spectra of GMP-selected systems, obtained with Keck/OSIRIS \citep{Larkin06} and HST/STIS, were presented by \cite{Mannucci22,Mannucci23} and \cite{Ciurlo23}.

Our aim is to discover and classify more GMP dual-AGN candidates. Here, we present AO-assisted observations with MUSE (Multi Unit Spectroscopic Explorer, \citealt{Bacon10}) narrow field mode (NFM) at ESO VLT (Very Large Telescope), of 12 GMP-selected systems at $z>0.3$, the largest such sample to date. 
We also present the spectral-decomposition analysis that we applied to spatially unresolved spectra to identify the AGN plus star systems, and to remove them from the target list of the AO observations.

The manuscript is structured as follows. In Sect.~\ref{sec:target}, we 
present the target selection and how an accurate analysis of the integrated spectra can help removing the AGN-star systems.
In Sect.~\ref{sec:obs} and ~\ref{sec:analysis},
we describe the observations and the steps taken to analyze the spectra and identify the nature of each component.
In Sect.~\ref{sec:results}, we analyze and classify each system.
Finally, in Sect.~\ref{sec:summary} we summarize our findings and discuss their implications in the larger context of observed dual AGN at small separations, comparing them with other results from the literature.

\begin{table*}[] \footnotesize
\caption[]{Main properties of the targets observed with VLT/MUSE.}
\label{tab:targets}
    \begin{tabular}{lccccccccccc}
\hline
 ~~~~~Target           & RA &DEC& Redshift  & \fmp & Gaia Sep.  & J &G1 &G2 & Spectrum  & T$_{\rm exp}$& Observation\\  
                  &                    &      &            &    & ($\arcsec$) &     & &  &  source&  (min:s)& date\\
\hline
J0007+0053  &00:07:10.02 & +00:53:29.0  & 0.318 &12  & 0.792 & 15.99 &17.68 &19.34& SDSS$^1$   & 19:06 & 11/20/22\\
J0015--2526$^*$  & 00:15:43.93   & --25:26:17.7     & 0.969  & 62 & -       & 15.27  & 16.27&-& NTT$^2$  &  19:06 & 12/20/22\\
J0052--3045$^*$  & 00:52:37.98  & --30:45:52.2   &1.462 & 44 & -        & 17.63 &19.78&- &2QZ$^3$   &  26:42  & 12/24/22  \\
J0128--0444 &  01:28:37.44 &  --04:44:01.5 & 1.631 & 15&- & 17.72 &20.42&-& SDSS$^1$    &   34:20  & 08/20/22\\
J0402--3237  &  04:02:15.06 & --32:37:33.4 & 1.280  & 16 & 0.661 &17.19& 19.47 &20.63& Lemon$^4$ &27:12 & 10/19/22\\
J0525--4153  & 05:25:35.88 & --41:53:46.1     & 2.155& 26  & -  & 16.95 &18.17&-& Gaia$^5$ & 26:42 &12/29/22\\
J0545--2236     &05:45:09.38  & --22:36:33.5    & 2.488 & 21& 0.554  & 17.31  & 19.51& 20.11 &Gaia$^5$& 26:42 & 01/15/23\\ 
J0840--0529    &  08:40:04.14 & --05:29:11.1    &1.393 &55 & -  & 17.01& 18.34&-& Gaia$^5$  & 26:42 & 01/25/23\\
J1059+0622$^*$   & 10:59:26.44 & +06:22:27.1   & 2.199 & 27 & 0.592 & 15.98 &17.47&17.95& SDSS$^1$   &   19:06 & 01/27/23 \\
J2139--4331   & 21:39:57.14  & --43:31:01.7  & 0.485 & 10 & 0.666   & 16.36 &18.57&18.68& NTT$^2$  & 19:06 & 11/20/22\\
J2211--0009      & 22:11:11.00 & --00:09:53.4     &  0.667 & 24 & -         & 17.17 &19.26&-& SDSS$^1$    &  34:20 & 07/03/22  \\
J2228--3047  &22:28:49.43  & --30:47:34.5  & 1.948 & 28  & -   & 16.87 & 18.19 &-& 2QZ$^3$ & 34:20 & 08/21/22\\
\hline
\end{tabular}
               \\

\footnotesize{
Notes: Separations (in case of multiple objects), G-band magnitude of the primary component (G1), G-band magnitude of the secondary component (G2, in case of multiple objects), and \fmp\ are extracted from the Gaia catalog; J-band are derived from Gaia Bp and Rp photometry, as is described in the appendix.  Redshift is obtained from spatially unresolved spectra whose origin is reported in the table.
T$_{exp}$ is the total exposure time of the MUSE observations.\\
$^*$ Targets on which we applied the decomposition method described in Sec. \ref{sec:deconv}.
$^1$: \cite{Lyke20}; 
$^2$: our observations at ESO/NTT (Program ID:109.22W4); 
$^3$: \cite{Croom09};
$^4$: \cite{Lemon20};
$^5$: Gaia quasars candidate catalog, \cite{Bailer-Jones23} 
}\\

\end{table*}

\begin{figure*}[]
  \centering
    \includegraphics[width=1 \textwidth ]{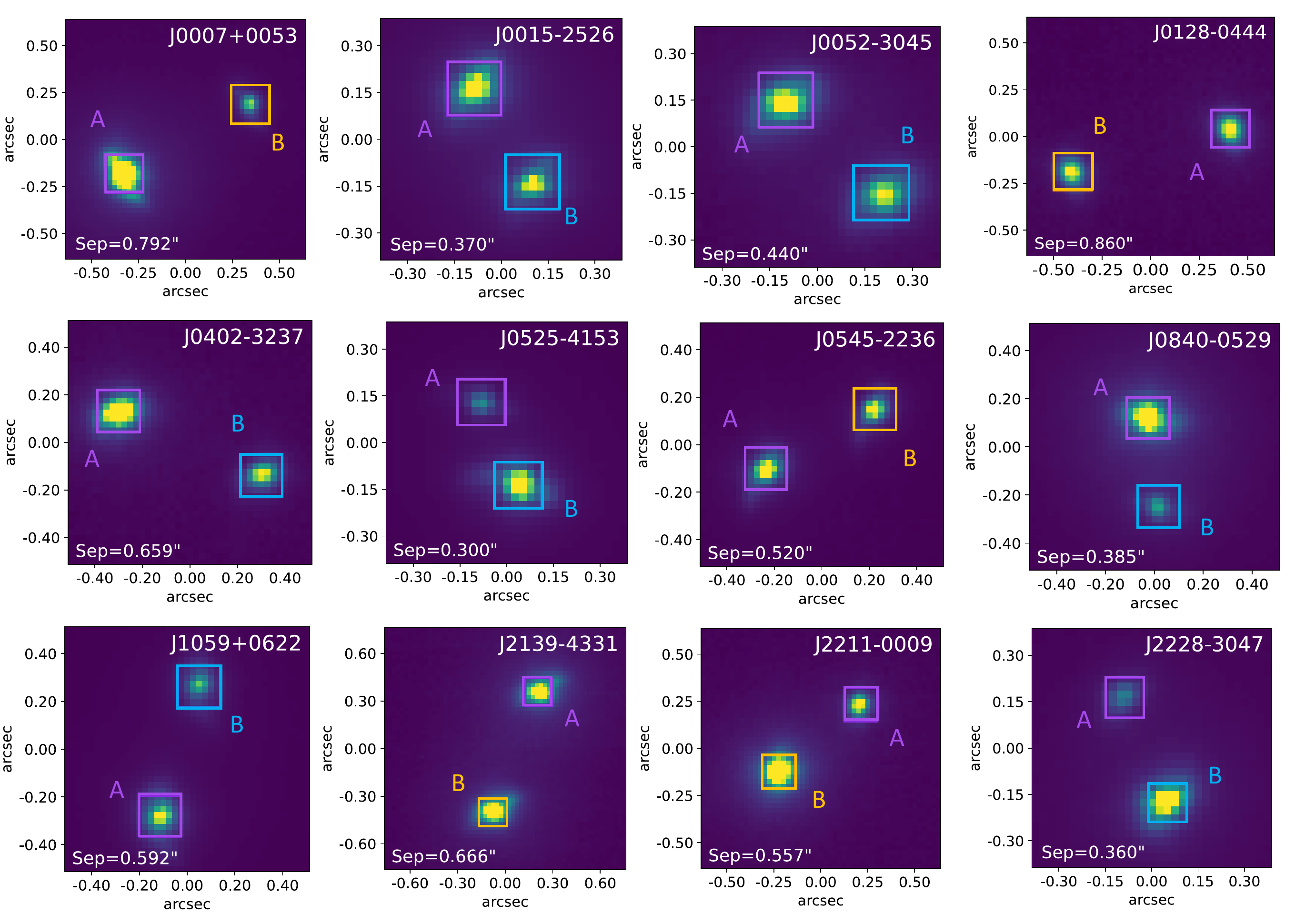}
    \caption{White light images (target name and separation of the two objects are reported on the figure). The spectra are extracted from the square apertures plotted over the maps. Orange squares enclose the targets later identified as stars, while the confirmed AGN are displayed with violet and light blue apertures (A and B components).
    }
    \label{fig:cubes}
\end{figure*}
 \section{Target selection }
 \label{sec:target}
We selected 12 targets observable from VLT from a starting sample of AGN with \fmp>8, to ensure the presence of a double source \citep{Mannucci22}.

Among these systems, we selected targets with the following requirements:  a declination accessible from Paranal (Dec$<+15^{\circ}$);  a far-enough distance from the galactic plane ($|b|>15^{\circ}$) to reduce stellar contamination; and a sufficient brightness in the near-IR to be used as tip-tilt (TT) star for the MUSE AO system ($J<17.5$). This threshold was chosen to be one magnitude brighter than the limit given in the MUSE manual for P109, to allow for AO observations in turbulence class up to $50\%$. $J$ band magnitudes were either obtained from 2MASS, or estimated from the observed Gaia Bp and Rp magnitudes with an uncertainty of $\sim 0.27$, as described in the appendix. Moreover, we chose targets far from large galaxies in the Local Group (mainly M31, Large and Small Magellanic Clouds) in order to decrease stellar contamination. Finally, we preferentially selected objects with  $z>0.5$ to probe the redshift range at which most of the merging activity was expected \citep{Chen23b}, and to avoid the presence of bright host galaxies that could affect the GMP selection \citep{Mannucci22}.

Nine of the targets were selected among spectroscopically confirmed AGN, ; that is, objects for which previous ground-based, seeing-limited spectroscopy is available (\S~\ref{sec:spec_targets}). \\
We complemented this sample with 3 additional photometric GMP-selected quasars 
(see \S~\ref{sec:phot_targets}) from the AGN Gaia catalog for which only low-resolution spectroscopy (R$\sim$20$-$100) from Gaia is available.

The observations were carried out from July 2022 to March 2023. The main properties of these 12 targets are reported in Table~\ref{tab:targets}.

 \subsection{Spectroscopic targets}
 \label{sec:spec_targets}
We chose nine sources among $\sim 300$ GMP-selected targets with spectroscopic redshift from the Milliquas v$7.2$ catalog \citep{Flesch21} and from our own spectroscopic follow-up (Program ID:109.22W4, PI: Mannucci) obtained with the ESO Faint Object Spectrograph and Camera (v.2, EFOSC2) at the New Technology Telescope (NTT). The results of this spectroscopic campaign will be presented in a future publication.

The spectra of part of the sample were examined to exclude objects showing the presence of a star in their ground-based unresolved spectrum. To achieve this, we developed the novel procedure described in the next section.



\subsubsection{Decomposition of ground-based spectra}
\label{sec:deconv}

To identify the systems resulting from chance alignments between AGN and Galactic stars, we developed a procedure to examine the total combined spectra of unresolved multiple GMP-selected sources before the MUSE follow ups. This method enabled us to eliminate from the target list those systems showing significant stellar features at velocity $\sim 0$ km/s with respect to Earth, indication that the companion of the AGN is a star \citep[see also][]{Shen23b}.

Ground-based spectroscopic data (see Table~\ref{tab:targets}) covering both components of the pairs were primarily used to detect the presence of an AGN and to measure its redshift. 
In GMP-selected systems, the luminosity ratio between primary and secondary component is usually less than a factor of six for all the objects with $G>18.5$ \citep{Mannucci23}. Both components are expected to contribute significantly to the observed spectrum. As a consequence, we proceeded by decomposing the total ground-based spectrum into a combination of a quasar and a star with an innovative routine (see Appendix A in \citealt{Mannucci23} for the first application of this routine).

For the dust-obscured quasar component, we adopted the parametric spectral energy distribution (SED) model from \cite{Temple21} to reproduce the average observed quasar spectra across the redshift range $0<z<5$.  

For the stellar component, we used a library consisting of a combination of the MaStar (MaNGA Stellar Library) survey from SDSS-IV (see \citealt{Yan19}), covering the spectral range $3622 - 10354\ \AA$ at a resolving power of $R \sim 1800$. We only considered stars of spectral types G, K, and M, as hotter stars are very rare at high Galactic altitudes. To fit the observed spectra, we considered all the possible combinations of 30 stellar spectra and 6 emission-line QSO templates (see Appendix B in \citealt{Temple21} for details). Different QSO templates span from the minimum to the maximum of the range of observed QSO SEDs, accounting for different levels of ionization of emission-line regions.  The decomposition routine degrades the spectral resolution of the template spectra to be the same of the observed data and finds the linear combination of stellar and QSO templates that provides the best-fit to the observed spectrum.

The free parameters of our decomposition procedure include the stellar spectral class, the amount of dust extinction in the QSO spectrum, $A_V$, $E(B-V)$, and the redshift of the QSO, z$_{\rm QSO}$, if not previously known. Other free parameters include the flux normalization of the two components, $F_{\rm star}$ and $F_{\rm QSO}$, and the radial motion of the stellar companion, allowing for a maximum shift of the stellar spectrum of $\pm 300$ km/s 
\footnote{See \cite{Lindegren21} for a comparison of astrometric and spectroscopic nearby stellar motion measurements.} 
For each set of parameters we test every possible combination of stellar plus QSO templates, performing a loop over the parameter space until obtaining the set of free parameters that minimizes the residuals. Before comparing the modeled and observed spectra, we accounted for the spectral resolution of ground-based observations by degrading the templates resolution. This is necessary to properly compare emission and absorption features which play a key role in this routine. Since in many cases we observe the co-presence of strong QSO emission lines and weak stellar absorption lines, we determine whether to include a stellar component by estimating the $\chi^2$ over the main stellar absorption features for each spectral type.

With all these ingredients, we ran the spectral decomposition and classified a system as an alignment of an AGN and a star if the addition of a stellar component decreases the $\chi^2$ computed over the main absorption lines by at least $30 \%$. While fitting the data with a combination of QSO and stellar templates, and in order to reduce degeneracies between different spectra combination, we also imposed that the magnitudes of the primary and secondary systems obtained from the fit be within the sensitivity limits of the GMP method (see Fig.2 in \citealt{Mannucci23}). 
This decomposition requires a S/N and a resolution of the spectra sufficient to identify and characterize the features of the star. 
We expect this decomposition routine to be very effective in revealing the presence of a star if it has similar apparent luminosity of the primary QSO. If the luminosity ratio is close to the largest value allowed by the GMP selection ($\sim 2$ mag for targets with G$<$18.5), this procedure could be unable to detect the presence of a star. 

We developed this optimisation routine in November 2022, and used it since then to examine spectroscopic data of GMP sources for the presence of stellar features and define the MUSE target sample. 
At that date, six spectroscopic targets had been already observed (see \texttt{Observation date} in Table \ref{tab:targets}). Consequently, we applied this new decomposition method to the targets J0015--2526, J1059+0622, and J0052--3045.
As discussed in Sect.~\ref{sec:summary}, this procedure turned out to be very effective in identifying and removing AGN--star systems before spatially resolved observations. As a consequence, the chance to observe dual and lensed systems with IFU observations drastically increases.

\subsection{Non-spectroscopic Gaia AGN candidates}
 \label{sec:phot_targets}
We complemented the nine spectroscopic targets described above with 3 systems from the Gaia AGN catalog selected according to the procedure described below. We selected objects with values of \fmp\ above the adopted threshold of 8 (see Sec. \ref{sec:spec_targets}) among those classified as AGN candidates based on the Discrete Source Classifier (DSC). DSC is a module that classifies Gaia sources probabilistically into 5 classes: quasar, galaxy, star, white dwarf, and physical binary star. 
QSO candidates are identified by high values of the probability parameter \texttt{classprob\_dsc\_combmod\_quasar}, having values between 0 and 1.
We selected objects classified as very probable QSO (\texttt{classprob\_dsc\_combmod\_quasar} $>0.97$). 
This threshold value was chosen as a compromise between reliability of the classification and the number of observable targets.
We also required high values of two additional parameters provided by the Quasar Classifier (QSOC, \citealt{Delchambre22}) module. The first parameter,  \texttt{Zqsoc}, indicates the reliability of the redshift estimated by Gaia, while the processing binary flag, \texttt{flags\_qsoc}, reports the potential errors that can occur during the prediction process from QSOC. We considered sources with \texttt{Zqsoc} above the threshold of 0.99 (out of a maximum of 1) in order to obtain optimal redshift estimates and with the processing binary flag \texttt{flags\_qsoc}=0, indicating the absence of any warning flags.

\section{Observations and data reduction}
\label{sec:obs}

All the objects were observed with VLT/MUSE, which provides optical, integral-field unit (IFU) spectra (programs ID: 109.22W5 and 110.23SM, PI: Mannucci). This paper includes data acquired until March, 2023. The main properties of the 12 observed targets are reported in Table \ref{tab:targets}.
We used the $7.5\arcsec\times7.5\arcsec$ field of view (FoV) NFM of the instrument, assisted by the AO system (GALACSI, \citealt{Strobele12}) employing 4 laser guide stars (LGS, \citealt{Lewis13}). The spatially resolved spectra obtained have a spatial sampling of $0.025''$ $\times$ $0.025''$ and cover a spectral range from 4750 \AA\ to 9350 \AA\ with a resolving power ranging from 1700 in the blue to 3400 in the red.
In NFM-AO mode the range 5780$-$6050 \AA\ is blocked to avoid contamination by sodium light from the LGS (Na Notch filter).
We employed slightly different observing strategies in P109 and P110. For the former semester, we used 4 MUSE position angles (PAs) rotated by 90 deg from each other, for a total exposure time of 35~min per object. Having verified that 3 PAs are enough to obtain a good sky subtraction and remove the instrument artifacts, in P110 we reduced the overheads  by decreasing the PAs to 3, for a total exposure time between 19~min and 27~min per target, depending on the magnitude of each of them.
To ensure the highest quality observations and data accuracy, we require seeing full width at half maximum (FWHM) better than 0.8"
and a fraction of lunar illumination less than 50\%.

We reduced our data by using the processing pipeline for the MUSE Instrument \citep{Weilbacher20} running under the \texttt{EsoReflex} (ESO Recipe Flexible Execution Workbench, \citealt{Freudling13}) graphical environment. We used the standard workflow, subtracting the sky by extracting it from  $50\%$ faintest parts of the field of view, away from the target, which takes up only a small portion of the FoV.

In all cases, the target itself is used as reference star for the AO system. We didn't find problems in closing the AO loop on these double targets if the total magnitude is significantly brighter than the nominal MUSE limit of $J\sim18$, depending on seeing conditions. The AO system always provided an excellent correction, and we measure FWHMs of the point-spread-function (PSF) at $\sim7000\ \AA$ between 0.14" and 0.17".
J0052--3045 exhibits the worst PSF, with a FWHM $\simeq 0.17"$ in the direction along the separation. Despite this, since the separation among the components is 0.44" (see Tab. \ref{tab:muse}), the two sources are well distinguished even in this case.

\begin{figure*}
\centering
 \includegraphics[width=1\textwidth]{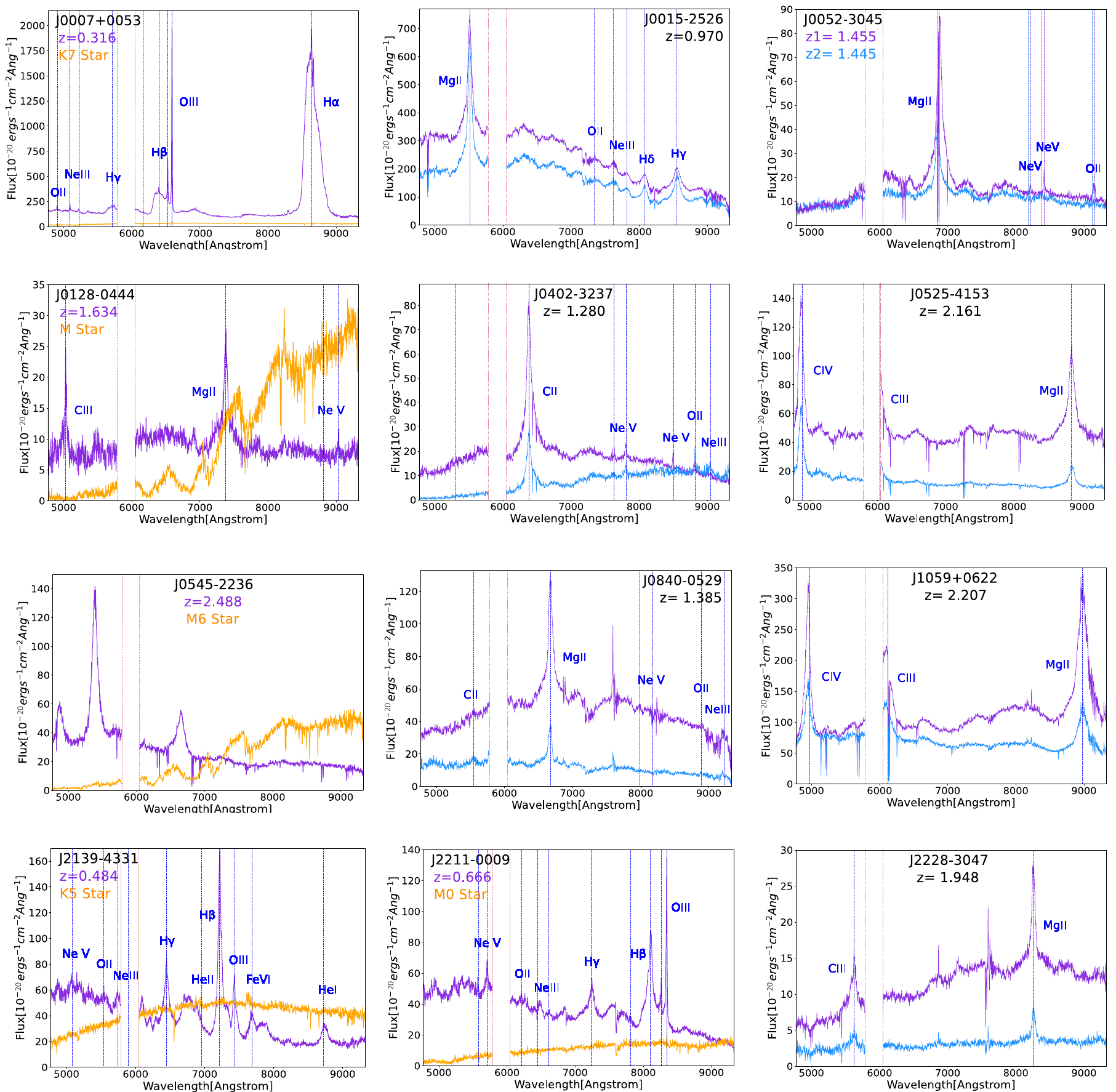}
    \caption{Spectra of the systems observed with MUSE (target name and redshift are reported in the panels). They have been extracted over the square apertures marked on the white light images in Figure \ref{fig:cubes} (with the same color-coding). Vertical dotted blue lines mark the position of the main expected emission lines. For J0052$-$3045, emission lines are plotted in light blue and violet as the redshift for the two AGN is not the same. The star in J0007$-$0053 has been multiplied by ten to optimize the visualization. The gap around $6000\AA$ is due to the New Generation Controllers (NGC) deployed in adaptive optics. 
    }
    \label{fig:spectra}
\end{figure*}

\begin{table*}
\caption[!h]{Results of MUSE analysis of systems with two AGN. Projected separations in kpc are computed at the redshift of the QSO.}
\begin{center}
\begin{tabular}{ccccccc}
\label{tab:muse}
Target  &   F(2600)$^*$ & Sep (") & Sep (kpc)   & Redshift  & $E(B-V)$ &  Classification\\  
\hline
J0015-2526\_A &   315.0 & \mr{0.370}  & \mr{-}   & 0.971 & 0.050 & \mr{Lensed AGN}\\
J0015-2526\_B & 193.6  && & 0.971 &  0.068\\
\hline
J0052-3045\_A     &17.6&\mr{0.440}  & \mr{3.82} & 1.455 & 0.156 & \mr{Dual AGN}\\
J0052-3045\_B     &13.2& && 1.445 & 0.152 \\
 \hline
J0402-3237\_A     &20.3& \mr{0.666}  & \mr{5.73} & 1.280 & 0.189& \mr{Dual AGN}\\
J0402-3237\_B     & 2.1& & & 1.281 & 0.861 \\
\hline
J0525-4153\_A      &44.7&  \mr{0.300}  & \mr{-} & 2.161 & 0.155 & \mr{Lensed AGN}\\
J0525-4153\_B      &9.4&   &  & 2.161 & 0.014\\
 \hline
J0840-0529\_A       &52.3&\mr{0.385}  & \mr{3.33}&1.386 & 0.130 & \mr{Dual AGN} \\
J0840-0529\_B      &14.4& & & 1.384 & 0.022 \\
 \hline
J1059+0622\_A      &118.1& \mr{0.579}  & \mr{4.91} & 2.209
& 0.197& \mr{Lensed AGN}\\
J1059+0622\_B     &61.3&&  & 2.210 & 0.124 \\
 \hline
J2228-3047\_A     &13.3 & \mr{0.377}  & \mr{3.25} &1.949 & 0.338& \mr{Dual AGN}\\
J2228-3047\_B      &3.1 & &  & 1.951 & 0.271\\
\hline
\end{tabular}
\\
\end{center}

\qquad \qquad \quad \quad \quad \footnotesize{$^*$ Continuum at 2600 \AA\ in $10^{-20}$ erg/cm$^2$/s/\AA}

\end{table*}

\section{Data analysis}
\label{sec:analysis}
MUSE data confirm that the GMP selection is able to successfully recover multiple systems: all the 12 observed systems are resolved into two point-sources with separations between 0.30" and 0.86", within the expected range according to \cite{Mannucci22}.
In Figure \ref{fig:cubes}, we show the cut-outs  extracted from the data cubes around the position of the targets, whilst in Figure~\ref{fig:spectra} we present the spatially resolved spectra extracted from the squared apertures shown on the images. In all cases, the separations are much larger than the PSF FWHM, therefore all the spectra are independent and can be used to separately study the nature of each source and measure its properties. The dimensions of the squared aperture could influence the data by introducing corrections related to the PSF. However, these effects remain consistent for both components. Hence, this does not impact our analysis. Note that, given the limited Strehl ratios (SR) of our observations, our spectra only contain part of the total flux of the objects, the same fraction for the two components of each system. Our results are not affected by this effect because they are all based on relative quantities.

We modeled each spectrum with either an AGN or a stellar spectrum. For AGN, we used the templates created by \cite{Vandenberk01} and by \cite{Temple21}. To account for the effects of interstellar dust extinction, we considered 3 different extinction laws to the AGN templates: \cite{Calzetti00}, \cite{Cardelli89}, and an empirical quasar extinction law presented in \cite{Temple21}. We fit all the observed spectra with these AGN templates, by varying E(B$-$V) and the source redshift. For each dual and lensed AGN, the best-fit values and classification are presented in Table \ref{tab:muse}. Specifically, according to the quality of the fit, we used the \cite{Cardelli89} extinction law to correct both spectra of J0402-3227, while for all other spectra we applied the empirical extinction law established by \cite{Temple21}. 
On the other hand, Galactic extinction has a limited impact due to the high Galactic latitudes of the targets ($|b|>21^{\circ}$), and therefore is negligible and ignored.

For stars, we used the same library as detailed in Sect. \ref{sec:deconv}, considering stars of G, K, and M spectral types.
These stars have much redder colors than AGN and, in most cases, their spectral features dominate the red part of the spectrum. Stars hotter than spectral type G are exceedingly rare far from the galactic plane; thus we exclude them. 
Our fitting procedure tests all the possible stars with different normalizations and selects the best one with the $\chi^2$ method. We report all the results in Table \ref{tab:starclassification}. \\

For the systems classified as composed by two AGN, our goal is to assess similarities and differences among the spectral features of the two images and determine whether they are associated to the same AGN lensed by a foreground galaxy, or to two distinct quasars at small separations.
In principle, the spectra of the two images of a lensed system are expected to be identical, except for a normalization factor, since gravitational lensing is an achromatic phenomenon. In practice, intrinsic variability, microlensing, and differential dust extinction can introduce differences in the observed spectra and mimic the presence of a dual system.
\begin{enumerate}
    \item Variability can introduce differences between the spectra when coupled with the different time delays between the two images. In our case, this is not expected to be a problem. In fact, the time delay between two images at sub-arcsecond separations is a few days at most \citep{Wambsganss98, Lemon23}, while at the luminosities probed in this work ($L_{\rm bol}=10^{46-47}$\ erg s$^{-1}$), the BLR is expected to extend for hundreds of light days \citep{Bentz13,Grier19,Prince23}, and the narrow-lines to be emitted over at least hundreds of parsecs. As a result, no significant variations in line profile or line ratios is expected due to this effect. However, small differences are expected in the continuum and, thus, in the emission broad line equivalent width (EW).

    \item While lensing is acromatic, microlensing, caused by compact objects (for instance, stars) within the lens galaxy,  can produce chromatic effects.  Microlenses differentially magnify parts of the quasar emission regions, leading to time- and wavelength-dependent changes in the flux ratios of the images \citep{Wambsganss06}.  Microlenses generally act on source regions at scales that are comparable to or smaller than a few micro-arcsec; that is,the size of their angular Einstein radius. Consequently, it is possible that both the quasar's continuum region, such as the accretion disc, and the broad line region will experience micro-lensing effects. Given that magnification varies in relation to source size, the power-law continuum emission of quasars will exhibit greater magnification at bluer wavelengths, emitted by more compact regions \citep[e.g.,,][]{mosquera2013,jimenezvicente2015}.  This also applies to the emission lines with higher ionization, which are expected to be subject to microlensing effects larger than those of the lower-ionization ones. Line shapes can also be affected, since different parts of the BLR could be differentially magnified/de-magnified. In contrast, the narrow lines, emitted over regions of hundreds or thousands of pc, are not subject to any microlensing effect. 
    Therefore, we have to define if the differences in the continuum shape, the discrepancies in line equivalent width and the deformations in the line profiles can be due to the possible combination of differential extinction and/or microlensing within the lensing galaxy or not. The analysis of this phenomenon is reported in Sect. \ref{sec:microlensing}.

    \item The light paths corresponding to the different images of a lensed system pass through different regions of the lens, and can intercept different amounts of gas and dust. As a consequence, the observed spectra could be characterized by different amount of dust extinction and different EWs of the absorption lines associated with the lens. This is addressed in Sect.\ref{sec:crosscorrelation}.
\end{enumerate}

Absorption lines also provide information on the nature of the system. If one lensed image exhibits one or more narrow absorption lines (NALs) at the quasar's redshift (intrinsic NALs), then the other lensed image should also display the same feature. This is because the intrinsic NALs reflect the gas composition within the quasar. 
However, differences in the EW of NALs may arise in the case of lensing, either due to microlensing or due to the geometry of the lens, with the line of sight of each image crossing different regions of the lens \citep[e.g.,][]{Chartas,Green06}.
In contrast, dual AGN can present different lines shapes, both for narrow and broad lines, and can easily have different absorption features, since they are separated by few kpcs.

In summary, to assess the nature of all the systems in our study, we jointly analyze the emitted spectra of both AGN, focusing on the following properties:
\begin{itemize}
    \item Redshift
    \item EWs of the detected lines, and line profile.
    \item Emission line flux ratios
    \item Presence of narrow absorption lines
    \item Continuum level and shape
\end{itemize}
If all these quantities are consistent within the uncertainty in the two spectra, we classify the system as a lensed QSO candidate. Otherwise, significant differences lead us to identify them as dual AGN. The number of quantities that show significant differences between the sources is directly proportional to our confidence in the dual nature of the system.

\subsection{Cross-correlation}
\label{sec:crosscorrelation}
In order to automatize the classification of the system (lensed or dual) through the comparison of the two spectra, we developed a cross-correlation approach. This method employs 3 distinct sets of free parameters, adapted to the characteristics of the examined system through $\chi^2$ minimization. 

The basic concept of our approach is to test if two AGN spectra can be produced by a lensed AGN. We thus modify one of the two AGN spectra in all the possible ways that are compatible with the effects produced by lensing, with the aim of  matching the spectrum of the other AGN, applying our modifications step-by-step as follows:
\begin{enumerate}
\item Initially, we cross-correlate the two spectra using two free parameters only, namely, a shift in redshift and a magnification factor in flux. 
In the case where the first parameter significantly deviates from zero, i.e., the two components show different redshifts, we classify the system  as dual. In the opposite case, if the shift is compatible with zero, if a perfect match (within the errors) is obtained, the systems is classified as lensed candidate, otherwise we proceed applying other modifications.
\item As a first refinement to the previous approach, we introduce an additional free parameter that accounts for the different normalization between the continuum and the lines that the two spectra display. This is possible in a lensed system, because the continuum can vary on shorter timescale as a consequence of the AGN intrinsic variability or of microlensing effects, while the narrow lines are expected to be stable. This is the case for J0015--2526 for example, where we find that the spectra of the two components can be matched by using two different normalization factors for the continuum and the emission lines, differing by 67\%. Consequently, this system is classified as a lens.
\item As a second refinement, we multiply the observed spectra for a second degree polynomial in order to account for possible differences in the extinction law between the two light paths. This can happen as the images that we observe come from the same source but pass through different regions of the lensing galaxy. We proceed with this further correction in the case of J0525--4153, where the two sources have different values of E(B-V), 0.155 and 0.014 respectively. 
\end{enumerate}

\subsection{Line-fitting procedure}
\label{sec:fitlines}
Another crucial step to classify the systems, distinguishing dual from lensed AGN, entails the analysis of the spectral lines.
To analyze the emission lines, we fit multiple Gaussian functions with different centers, amplitudes, and widths to each line in all observed sources. Gaussian fitting with two components (i.e., one narrow and one broad) was employed for broad lines, in addition to a power law for the continuum, while only one Gaussian was used for the narrow lines. For all the lines of each object, we measure the center and we quantify their EW. Central wavelengths and EW values are reported in Table \ref{tab:features}.

\subsection{Presence of the lens} 
\label{sec:lens_detection}
For all systems constituted by two AGN, we further explored the possibility of identifying the lensing galaxy in our datacubes, as expected in case of a lensed system.

First, we examined the broad-band image obtained by collapsing the cube along the wavelength axis to look for the presence of a foreground galaxy. To increase the image sensitivity, we also produced separate images in 4 different wavelength ranges (5300, 6500, 7100, 8700)\AA\ devoid of bright sky. No lens galaxy is directly detected in any of these images of the systems analyzed in this work.

Secondly, to identify potential faint or dust-obscured galaxies hidden by the presence of bright images of the lensed QSO, we subtracted the contribution of the AGN and examined the residuals. We fitted the PSFs by summing two 2D elliptical Gaussian functions, and proceed to subtract the contribution of the two AGN components.  
In all our systems, the observed residuals can be attributed to an imperfect fit to the PSF, and no lensing galaxy is unambiguously detected.

A lens non-detection is not a smoking gun to discard the lensed AGN nature if the lens can be fainter than the detection limit. Thus, we investigate whether the luminosity upper limits of the undetected lens can be used to exclude its presence and therefore secure the dual AGN nature.

We can roughly estimate the expected luminosity of the lensing galaxy as a function of its unknown redshift. We use half of the observed separation as a proxy for the Einstein radius and calculate the mass of the lens galaxy inside it as a function of redshift ($M_{\star}(z)$) assuming a point-like mass distribution (e.g., eq. 21 in \citealt{Narayan96}).
We assume that the lenses are early-type galaxies; in other words, the faintest systems for a given mass, as they have the oldest stellar populations. Assuming that stars constitute the majority of the mass inside the Einstein radius and that they follow a Sersic profile with an index of $n=4$, by using the mass-size relation in \cite{vanderWel14} we determine the needed stellar mass as a function of redshift. From this, we use \cite{Longhetti09} to  estimate the K- and V-band magnitudes of the possible lensing galaxy as a function of redshift.
In every scenario, the possible lenses are brighter if either at low redshift (low mass, but nearby galaxy) or high redshift (close to the source, distant but very massive galaxies), and fainter for intermediate redshifts. \\ The resulting galaxies are then convolved with the PSF and added to the datacubes and residual maps.

In all cases except J0402--3237, the lenses are expected to be below the detection threshold at any redshift. In J0402--3237, characterized by the largest separation of 0.67" and therefore requiring the largest masses, the lens should be detected for any redshift, and its absence can aid us in classifying the system as dual rather than lensed.

In conclusion, for all the  systems except J0402--3237, the absence of a visible lens galaxy in MUSE data does not preclude the presence of a lensing effect. The sensitivity of all spectroscopic data may not be sufficient to reveal extended objects with low surface brightness and no bright emission lines, such as the lensing galaxy. This limitation persists even with the subtraction of the AGN PSF. The intricate structure of the PSF, influenced by atmospheric conditions, remains the primary factor hindering a clear delineation of the lens.

\subsection{Gravitational microlensing} 
\label{sec:microlensing}

Several techniques can be used to unveil microlensing in lensed quasars. In this work, we applied the so-called Macro-micro decomposition (MmD) from \cite{Sluse07}, which is based on comparing the differences in the (microlensed) continuum and broad lines to those in the (un-affected) narrow lines, once the macrolensing magnification is already accounted for. This method enables an easy visualization of the differential microlensing affecting the QSO spectrum. 

As stated by \cite{Sluse07} and \cite{Hutsemékers10}, we assume that the spectrum of an observed lensed image, denoted as $F_i(\lambda)$, can be decomposed into two distinct components. The first component, $F_M$, exclusively accounts for macro-lensing effects, while the second, $F_{M\mu}$, incorporates both macro and micro-lensing influences. We also defined the macro-magnification ratio between image 2 and image 1 as $M = M_2/M_1 (>0)$, the relative micro-lensing factor between the images as $\mu$, and $A$ is the scaling factor between the $ F_2$ and $ F_1$ continua. \\ 
The behaviors of $\mu$ and $M$ exhibit distinct characteristics. The factor $M$ may vary with wavelength due to differential extinction in the lensing galaxy. On the other hand, $\mu$ is presumed to remain constant within the limited wavelength range covered by a line profile, but chromatic microlensing may occur due to the wavelength dependence of the source size; the blue continuum-emitting region may be smaller and more microlensed than the red one. \\
Hence, we calculated $\mu$ and $M$ in close proximity to each emission line for every system.

To definitively assess the presence or absence of the microlensing effect, one needs to consider two lensed images and determine both components, $F_M$ and $F_{M\mu}$, using the following equations:
$$ F_1=M \times F_M +M\times\mu \times F_{M\mu} $$
$$
F_2=F_M + F_{M\mu} $$
Up to a scaling factor, $F_M$ only depends on A, while $F_{M\mu}$ depends solely on M. The A parameter can be accurately determined as the value for which $F_M(A)=0$ in the continuum adjacent to the line profile. Assuming A remains constant along the line profile, $F_M$ exclusively reflects the contribution from the emission profile. If we assume that a part of the observed emission experiences only macro-amplification (without micro-amplification), we can estimate the macro-amplification factor $M$. Given the relation $A = M \times \mu$, we can then estimate the micro-amplification factor $\mu$ for the continuum at the wavelength of the line profile.  On the other hand, if the emission line is amplified exactly like the continuum, microlensing cannot be distinguished from macrolensing.


\section{Results}
\label{sec:results}
The details of the resulting classification are listed in Table~\ref{tab:starclassification} for the 5 AGN/star associations, and in Table~\ref{tab:muse} for the seven dual AGN or lensed QSO candidates.
Following the cross-correlation procedure described in Sect. \ref{sec:analysis}, we classify 3 objects as lensed AGN candidates and 4 as dual systems. 

We classify J0015--2526, J0525--4153, and J1059+0622 as lensed AGN candidates since their components exhibit overlapping spectra, with similar trends in continuum and EW for all the lines.

On the other hand, J0052--3045 is a clear dual system because its spectra show a significant difference in
the lines on the NLR and in redshift, estimated to be $\Delta z=0.010$ ($\sim3000\ $ km/s) through cross correlation analysis. However, this value could be smaller due to the absence of narrow lines in one of the AGN (light blue in Fig. \ref{fig:spectra}) making it difficult to accurately estimate the redshift. As a result, its measurement relies solely on the broad narrow line MgII, which exhibits a distinctly different profile between the two components. In the other cases (J0402--3237, J0840--0529, and J2228--3047), the spectra reveal different EWs and line ratios. In J0402--3237 and J2228--3047, the continuum also differs. Additionally, in the latter object, the largest difference is observed in the EW of the iron lines, specifically in the region around 7000 \AA.

We note that some spectra reveal the presence of NALs from intervening material.
Here, we only discuss the absorption features of J0052-3045, which are used for the classification of the system. The other absorption systems will be described in a future work.

In the following, we describe the properties of each of the observed systems and the procedure followed for their final classification.

\subsection{AGN+star}
\label{sec:agnstar}

5 of the observed systems are associations of an AGN and a foreground star (see Table~\ref{tab:starclassification}). Figure \ref{fig:spectra} presents the MUSE spectra of these 5 systems, where the orange line shows the spectrum of the star. The classification of the secondary system as a star is performed through the technique presented in the previous section; that is, by finding the best-fitting stellar spectrum. Two of these systems, J0007+0053 and J2139--4331, have previously been observed and studied by \cite{Lemon20}. In that work,
J2139--4331 was already classified as an AGN+star system, a classification confirmed by our spectra, while no classification was provided for J0007+0053 due to the limited (seeing-limited) spatial resolution.\\
Moreover, it is noteworthy that among these systems, the separation of J2139--4331 measured by MUSE data (0.82$''$) differs significantly from the Gaia measurement (0.666$''$). This disparity can be attributed to the high proper motion of the star measured by Gaia.

\begin{table}[!h]
\caption[]{List of AGN+star systems observed in the sample, with their angular separation and best-fitting stellar type.}
\label{tab:starclassification}
\centering
\begin{tabular}{ccccc}
\hline
 Target          &   Sep (")    &Star Type  \\
\hline
J0007$+$0053 &  0.78  &  K7 \\
J0128$-$0444 &  0.86  &  M7 \\
J0545$-$2236 &  0.52  &  M6 \\
J2139$-$4331 &  0.82  &  K5 \\
J2211$-$0009 &  0.56  &  M0 \\
\hline
\end{tabular}
\\
\end{table}
\subsection{Lensed AGN systems}
\label{sec:lensedsystems}
Both the components of the seven remaining systems show AGN spectra.
We start by describing the results of the MUSE spectral analysis for the 3 systems classified as lensed AGN candidates.
The values of all the spectral features described below are reported in Table \ref{tab:features}.

\subsubsection{J0015$-$2526} 
\label{sec:J0015}
\begin{figure}[!h]
\centering    \includegraphics[width=0.5\textwidth]{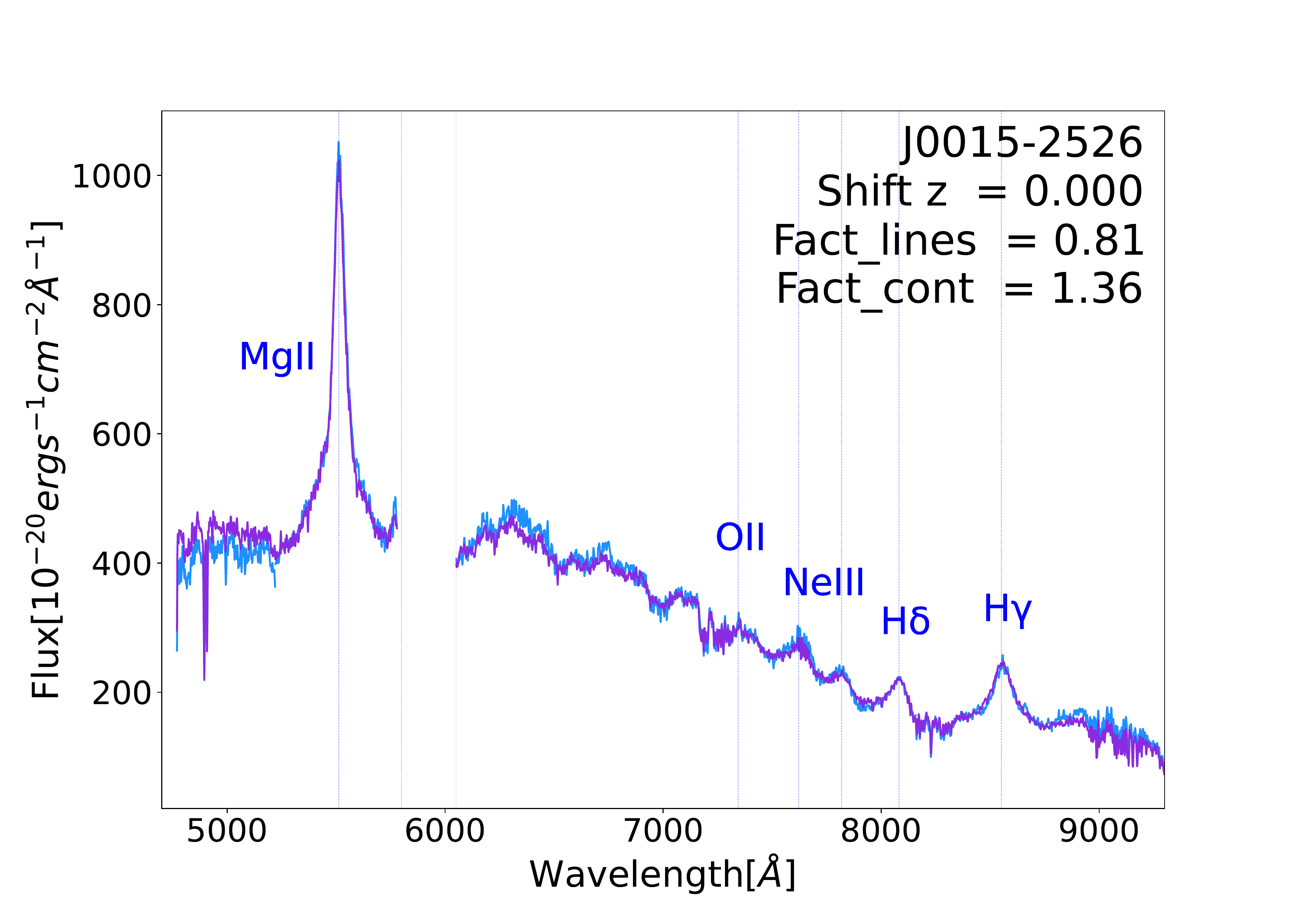}
    \caption{Cross correlation J0015$-$2526\_B (light blue AGN spectra): multiplied in flux and shifted to the same redshift by the quantities indicated in the labels. Color-coding as in Fig. \ref{fig:spectra}.}
    \label{fig:crossJ0015}
\end{figure}

The classification of this system as a lensed AGN candidate is based on the striking similarity of the two spectra, and confirmed through the cross-correlation discussed in Sect. \ref{sec:crosscorrelation} and shown in Fig. \ref{fig:crossJ0015}. For J0015$-$2526, we quantified a magnification factor for the emission lines (\texttt{Fact\_lines}) and applied a correction to this value to reproduce the continuum (\texttt{Fact\_cont}). The use of two different values is needed because the continuum could be affected by microlensing and could show variability on a short timescale \citep[e.g.,][]{Fian18}. The correction factor obtained, $67\%$, is in agreement with this scenario. The values of these two scaling factors are reported  in Fig.~\ref{fig:crossJ0015}.  

In the spectra of J0015$-$2526 there are six detected lines, of which MgII$\lambda2798$, H$\delta\ \lambda4103$, and H$\gamma\ \lambda 4342$ were fitted following the procedure explained in Sect. \ref{sec:fitlines}. 
Gaussian fitting with two components (i.e., one narrow and one broad) was employed for MgII, in addition to a power law for the continuum, while only one Gaussian was considered to model the Balmer lines due to the lower S/N. Central wavelength and EW values were computed for each line, and were found to be consistent between the two AGN. Additionally, the two sources have the same emission line flux ratios within the errors, providing strong evidence that this is indeed a lensed system. Notably, the $H_{\delta}/H_{\gamma}$ ratio, which is determined from two narrow lines, is perfectly equal in the two components.

Since J0015$-$2526 is classified as a lensed system, we tried to identify the lens as we explained in Sect. \ref{sec:lens_detection}. \\
First, we produced 4 images in 4 wavelength ranges devoid of bright sky lines, i.e.: $\lambda=(5300, 6500, 7100, 8700)~\AA$. We summed these images to search for any possible lensing galaxy in the field. The lens galaxy is not detected, in either the summed image or the partial images. \\
Second, we investigated the possible properties of a lensing galaxy to test if its non-detection in the MUSE datacube is consistent with the lensed AGN nature of J0015$-$2526. Based on the measured quasar redshift ($z=0.970$). We tested this for 3 lens redshifts: 0.25, 0.50, 0.75. For each redshift value, as explained in \ref{sec:lens_detection}, we derived the mass enclosed by the Einstein ring, and used \cite{vanderWel14} and \cite{Longhetti09} to assess the size and magnitude. We then simulated the 3 galaxies assuming a Sersic profile with n=4 and we add each galaxy to our data. With these magnitudes (G > 21.8, 22.5, 22.0, respectively), none of the tested lensing galaxies is brighter than the limiting sensitivity of our data ($G\sim20$, estimated by adding compact galaxies of various brightness to the data). Furthermore, we observe some absorption lines in the spectra of J0015 that are indicative of an intervening absorber at z=0.75. The lens corresponding to this redshift could be a typical elliptical with mass $M=10^{11} M_{\odot}$, and an effective radius $R_{eff}=0.4 \arcsec$, with a magnitude of $G=22.4$, thus undetectable in our data.
\begin{table}[]
\caption[]{Summary of the observed spectral features discussed in sections \ref{sec:lensedsystems} and \ref{sec:dualsystems}. EWs are reported in \AA.}
\label{tab:features}
\centering
\begin{tabular}{rcc}
\hline
Feature & Comp. A & Comp. B \\
\hline
        & \mc{J0015-2526}\\
\hline
EW(MgII) &  $56\pm 6$ & $111\pm 12$ \\
EW($H\delta)$ &   $24\pm4$ & $33\pm7$ \\
EW($H\gamma)$ & $64 \pm 6$ & $97\pm7$\\
MgII/$H\gamma$& $2.1\pm0.5$ &  $2.7\pm0.6$\\
MgII/$H\delta$& $4.6\pm1.0$ &  $6.8\pm1.5$\\
$H\gamma/H\delta$ & $2.2\pm0.5$ & $2.5\pm0.6$\\
$\lambda_0$(MgII) & $5513.9 \pm 0.3$ & $5514.7\pm0.2$ \\

\hline
      &  \mc{J0052-3045}  \\
\hline
EW(MgII)          & $355 \pm 40$ & $174 \pm 20$\\
$\lambda_0$(MgII) & $6890.2 \pm 1.3$ &   $6855.3 \pm 0.5$\\     
\hline
          & \mc{J0402-3237} \\
\hline
EW(MgII)  & $128 \pm 13 $ & $326 \pm 37$ \\
EW([NeV])   & $11.9\pm0.9$   & $15 \pm 6$ \\
EW([NeIII]) & $8.9 \pm 1.3$ & $13.5 \pm 1.2$ \\
MgII/[NeV]  & $14 \pm 1$      & $29 \pm 6$ \\
MgII/[NeIII]  & $29 \pm 7 $      & $41 \pm 9$ \\
$[$NeV$]$/[NeIII] & $2.1\pm 0.5$      & $1.4 \pm 0.3$ \\
$\lambda_0$(MgII) & $6381.0 \pm 0.3$ & $6381.1\pm0.3$ \\

\hline 
     &    \mc{J0525-4153} \\
\hline 
EW(MgII) &   $83 \pm  10$ & $102 \pm  12$ \\
EW(CIV)& $60 \pm 18$ & $175 \pm 5$ \\
MgII/CIV& $0.24 \pm 0.05$ & $0.29 \pm 0.06$ \\
$\lambda_0$(MgII) & $8851.1 \pm 0.4$ & $8856.5\pm0.4$ \\
$\lambda_0$(CIV) &  $4878.4 \pm 0.3$ & $4877.9 \pm 0.3$ \\
\hline
                 &  \mc{J0840-0529}  \\
\hline
EW(MgII)         & $24 \pm 3 $ & $41 \pm 4$ \\
$\lambda_0$(MgII)& $6677.0\pm0.4$    & $6670.8 \pm 0.4$ \\

\hline
                 &  \mc{J1059+0622} \\
\hline
EW(MgII)         & $231 \pm 26$ & $174 \pm 20$ \\
EW(CIV)          & $230 \pm 26$ & $118 \pm 3$ \\
MgII/CIV         & $0.48 \pm 0.11$ & $ 0.64 \pm 0.15$ \\
$\lambda_0$(MgII)& $8979.7\pm0.8$    & $8983.3 \pm 1.2$ \\
$\lambda_0$(CIV) & $4946.4 \pm 0.4$ & $4948.0 \pm 0.9$\\
\hline
          &  \mc{J2228-3047} \\ 
\hline
EW(MgII)         &   $46\pm 5$& $63\pm 7$ \\
EW(CIII])         & $132 \pm 14$& $110 \pm2$ \\
MgII/CIII]     & $0.24\pm 0.05$ & $0.38\pm0.09$ \\
$\lambda_0$(MgII)& $8258.2 \pm 0.3$ & $8262.8 \pm 0.5$ \\ 
$\lambda_0$(CIII])& $5617.1 \pm 1.2$ & $5620.0 \pm 1.9$ \\
\hline
\end{tabular}
\end{table}

\subsubsection{J0525$-$4153}
\begin{figure}[!h]
\centering
    
    \includegraphics[width=0.5\textwidth]{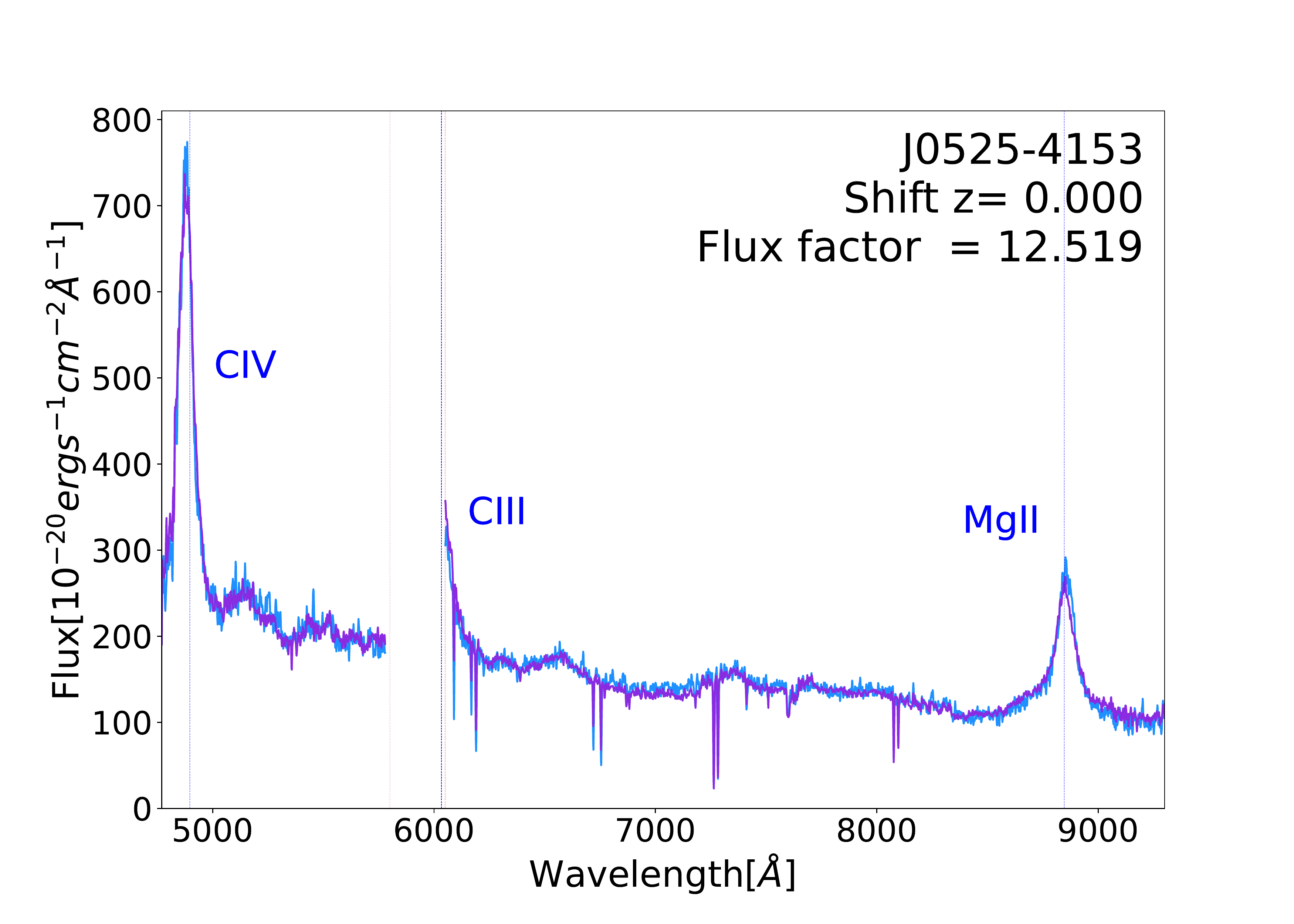}
    \caption{Cross correlation J0525$-$4153\_B (light blue AGN spectra): multiplied in flux and shifted to the same redshift by the quantities indicated in the labels. Color-coding as in Fig. \ref{fig:spectra}.}
    \label{fig:crossJ0525}
\end{figure}

The AGN J0525$-$4153, at z=2.155, is the first observed object selected from the Gaia catalog without a previous spectroscopic confirmation. The cross correlation, obtained by also considering a second degree polynomial for a supplementary difference in the extinction, clearly shows a close match between the spectra (Fig. \ref{fig:crossJ0525}). Continuum and lines are perfectly overlapping, as well as the several narrow absorption lines, strongly suggesting that this is a lensed AGN.

To study the emission lines, we primarily focus on the MgII and CIV emission line components, which we analyzed through the Gaussian fitting procedure.
For the MgII emission line, we employed two Gaussian profiles and a power-law component, while in the case of the CIV line we made a fit with a Gaussian and a power-law profile only. 
We estimated the equivalent widths of both lines, which were found to be compatible, yielding mutually consistent line ratios (MgII/CIV). 
Moreover, we observed that CIV is blue-shifted by the same velocity in both spectra, providing further confirmation of the lensed nature of this system.

Additionally, we identified the presence of intervening NALs at the same observed wavelength in both spectra, suggesting that they originate from the same gas along the line of sight of the quasar. 
We identify the observed lines as FeII$\lambda2382$, FeII$\lambda2585$, FeII$\lambda2600$, MgII$\lambda2796$ and MgII$\lambda2803$ at z=1.597 and z=1.889. These intervening NALs could be due to the lens galaxy located between the two AGN images, or they could be associated with other objects along the line of sight.. Further investigation of the NALs could yield even more detailed information about the lens galaxy and other intervening objects, their redshift and chemical abundances. 
However, as in the previous case, our data are not deep enough to detect the lens in a system with this configuration.

In summary, our analysis of the spectra of J0525$-$4153 has provided strong evidence supporting its identification as a lensed AGN candidate. 


\subsubsection{J1059+0622}
J1059+0622 was previously analyzed by \cite{Lemon17} who classified it as a lensed candidate based on variability.

The spectra of J1059+0622 exhibit the presence of 3 emission lines, although one of them (CIII]) is partially lost due to the LGS gap (see Fig. \ref{fig:spectra}).
Since the spectra seem to be apparently similar as is the case for gravitationally lensed sources, we tried all the cross-correlation types to test this scenario. However, none of the tested spectral modifications produce perfectly overlapping spectra and a clear difference in CIV line is always present. 

\begin{figure}[]
\centering
    \includegraphics[width=0.4\textwidth]{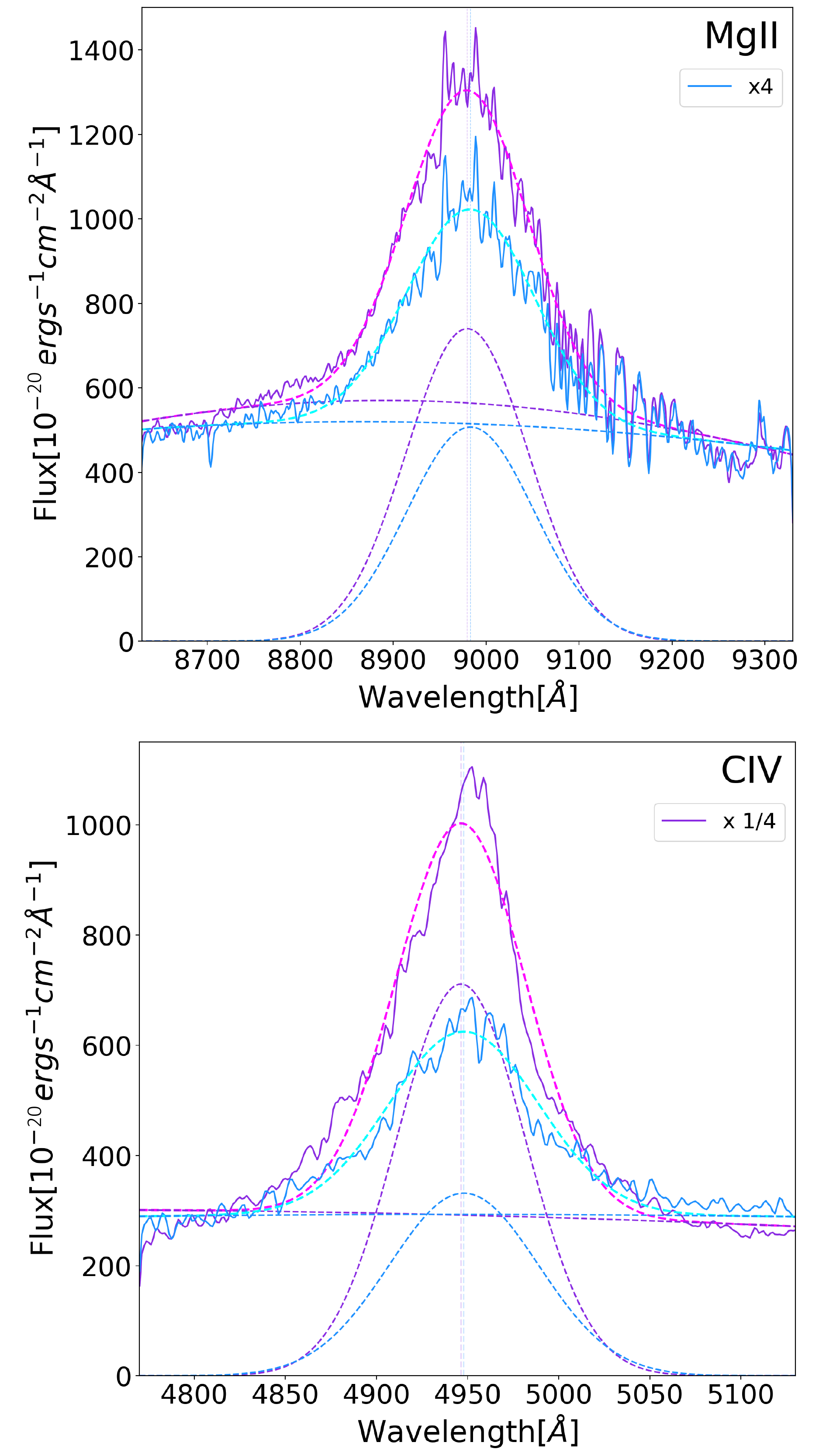}
    \caption{Fit to the MgII (top) and CIV (bottom) lines of both components of J1059+0622. Color-coding of Fig \ref{fig:spectra}. Each component of the fit is shown with dotted violet and light blue line, for components A and B, respectively. }
    \label{fig:fitLinesJ1059}
    \end{figure}

\begin{figure}[]
\centering
       \includegraphics[width=0.5\textwidth]{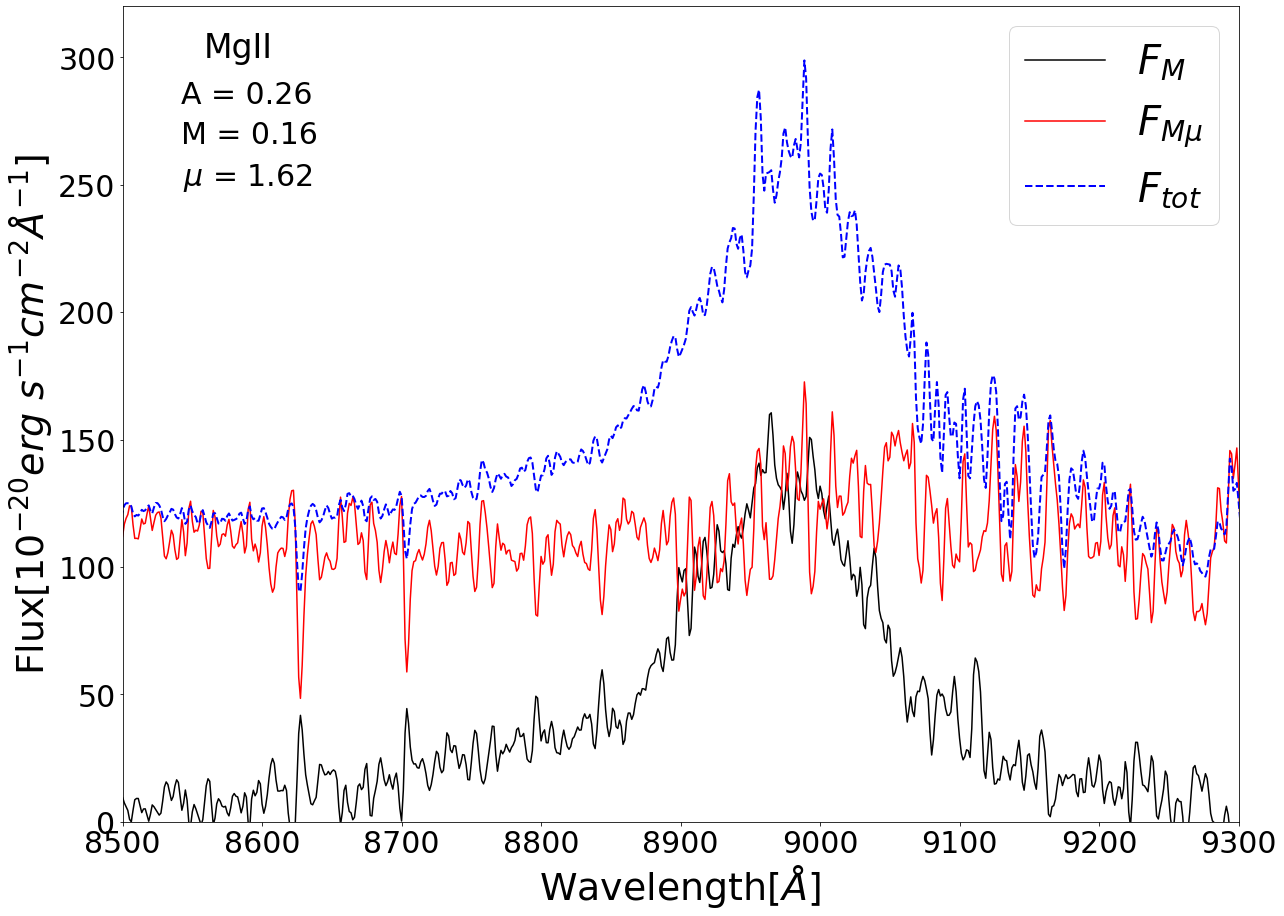}
       \includegraphics[width=0.5\textwidth]{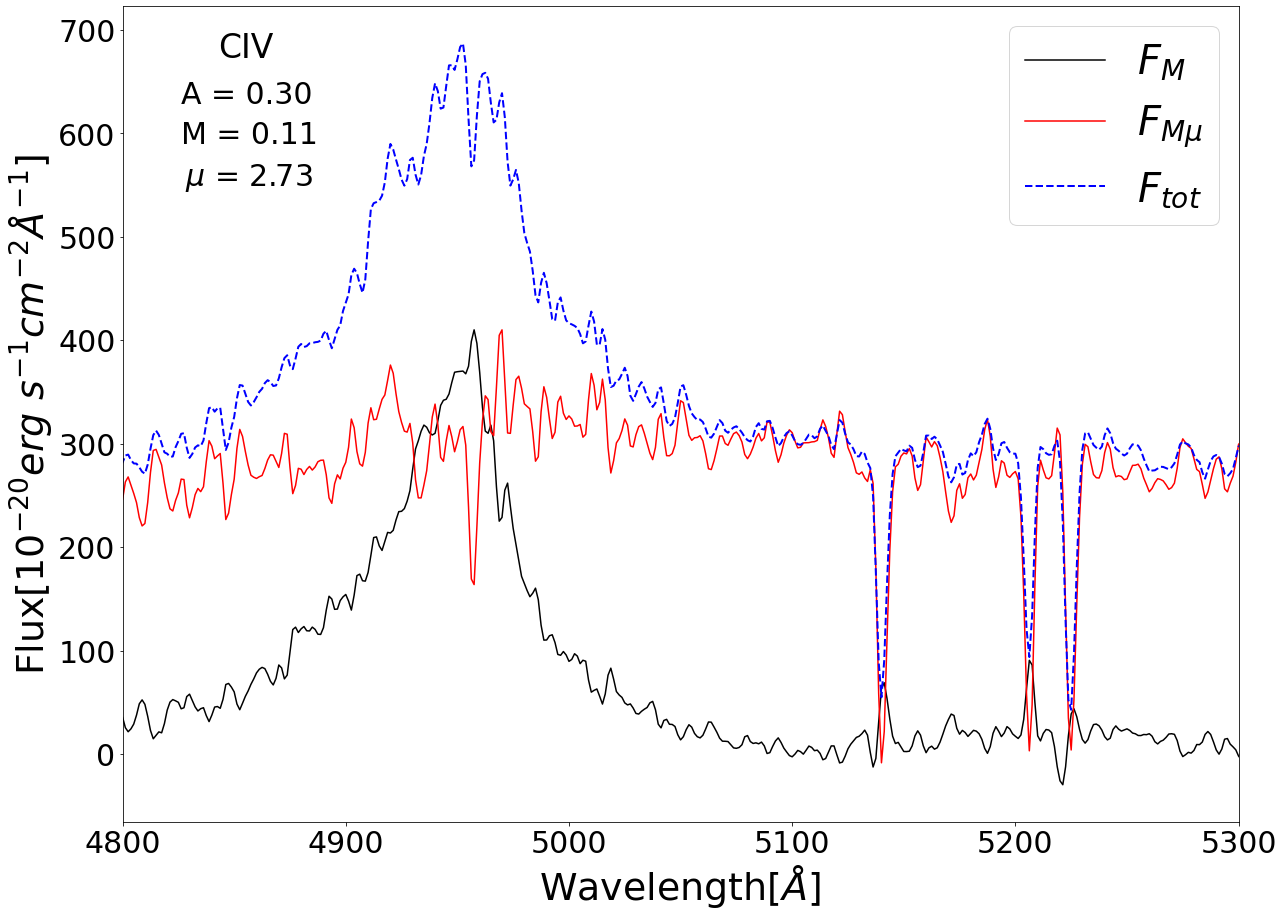}

    \caption{MmD decomposition following the procedure in \cite{Sluse07} on the dereddened spectra of J1059+0622. The upper panel shows the CIV line, while the lower panel displays the MgII line. The resulting values of the relevant parameters are reported in the figure
 }
    \label{fig:microJ1059}
    \end{figure}

We fit MgII and CIV lines with a Gaussian component plus a power-law for the continuum (dotted violet and light blue lines in Fig. \ref{fig:fitLinesJ1059}). 
The estimated EWs and 
line ratio MgII/CIV for the two components differ significantly. We report the obtained values in Table \ref{tab:features}. 
Since these differences can be caused by microlensing, we tested the presence of this effect with the MmD procedure of \cite{Sluse07} (described in Sect. \ref{sec:microlensing}).
Applying it to both the observed spectra and the dereddened ones, we identified a plausible scenario involving microlensing of the continuum along with differential extinction, potentially attributable to a lensing galaxy. 
The derived MmD values for the CIV and MgII in the observed spectra are as follows: $A=1.0,\ 0.53$, $M=0.38,\ 0.32$, and $\mu=2.63,\ 1.66$, respectively. 
The difference of the value of M for the two lines is compatible with the differential extinction $\Delta(E(B-V))=0.73$ reported in Table~\ref{tab:muse}, thus these values are perfectly compatible with differential extinction plus microlensing of the continuum \citep{Sluse12}. The same interpretation holds for the dereddened spectra, for which we obtain the following MmD values for the CIV and MgII lines: $A=0.3,\ 0.11$, $M=0.26,\ 0.16$, and $\mu=2.73,\ 1.62$ (Fig. \ref{fig:microJ1059}). Also in the dereddened spectra, the weak deformation of MgII can be due to residual differential extinction. This might be related to the empirical extinction law applied during the dereddening process (\cite{Temple21}, Sec. \ref{sec:analysis}), however, we evaluated the robustness of the decomposition with the "pre-processing" of the spectra by applying the decomposition on the dereddened spectra.
The discrepancy in E(B-V) could be ascribed to differential reddening of the two AGN images due to the two distinct lines of sight that probe different regions of the lensing galaxy. Therefore, the spectra are consistent with coming from two images of a single, lensed source.

Another indicative feature that prompts consideration of a lensed system is the presence of intervening NALs at a redshift of 1.19 in both spectra. These may potentially be generated by the lens at this particular redshift.
However, there is no hint for residual contamination by a potential lensing galaxy.
Applying the procedure outlined in section \ref{sec:lens_detection} similarly to what done for J0015, a typical lensing galaxy at $z=1.19$ would not be detected in our data and neither would be if placed at a different redshift, given that the limiting magnitude of the MUSE observation is G$\sim20.5$. Thus, a lens non-detection is not a good enough reason to discard the lensing scenario for this source.

In summary, the most plausible classification for this system is a lensed AGN. Variations in the equivalent widths, preventing a perfect cross-correlation, are likely a result of microlensing and differential extinction.


\subsection{Dual AGN systems}
\label{sec:dualsystems}
In the following, we present the results of our MUSE data analysis of the 4 systems classified as reliable dual AGN.

\begin{figure}[]
    \centering
    \includegraphics[width=0.45\textwidth]{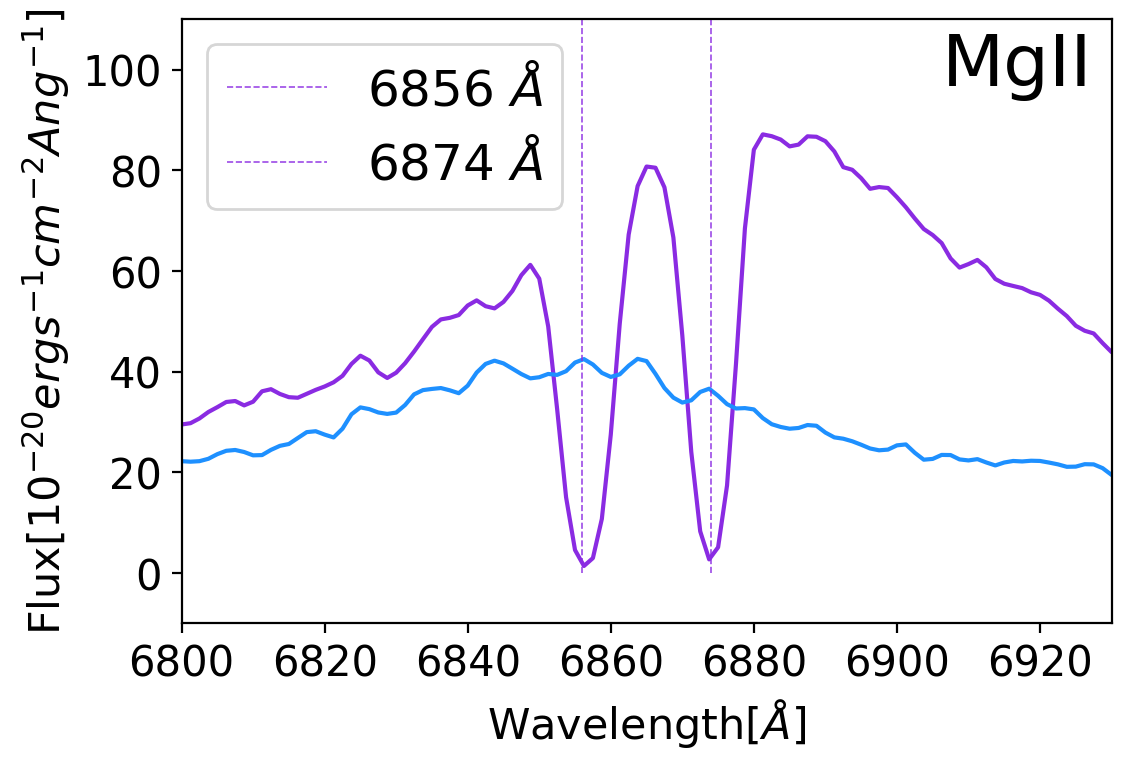}
    \caption{Enlargement of the MgII doublet absorption lines of J0052$-$3045. The lines are found at rest-frame wavelengths of  2796 \AA\ and 2803 \AA. Color code is the same of Fig. \ref{fig:spectra}.}
    \label{fig:zoomabsJ0052}
    \end{figure}
\subsubsection{J0052$-$3045}
J0052$-$3045 stands out among the analyzed AGN as it exhibits clear differences in redshift, line position, and line profiles, indicating that this is a genuine dual system (see Fig. \ref{fig:spectra}). The MgII line is present in both spectra, with a different shape and a velocity shift of approximately $\simeq1000$~km/s, corresponding to redshifts of $z=1.455$ and $z=1.445$ for the two components (see Table~\ref{tab:features}). This shift cannot be attributed to variability in the line profile since the expected time delay between different lensed images is a few days at most. 
Furthermore, shape and line ratios of narrow emission lines (i.e. [NeV] and [OII]) are very different.  These differences cannot be easily explained by the variability of the quasar, hence they provide an additional evidence of the distinct nature of the two AGN.  

Moreover, as we can see in Fig. \ref{fig:zoomabsJ0052}, there are large differences in the visible intrinsic absorption lines in the two distinct spectra.  The A component (violet) shows the MgII doublet in absorption at rest-frame wavelengths of $2796$ \AA\ and $2803$ \AA, ; in other words, at the same  redshift of the QSO. 
Since these NALs are entirely absent in the second AGN, gas composition of the two AGN is different, providing evidence that they cannot be two images of the same source.
The presence of these lines in only one of the two AGN further supports the hypothesis of a dual system, as one AGN has an outflow that is not visible in the other. According to all our findings, J0052$-$3045 is a dual AGN.

\begin{figure*}[!h]
    \centering
    \includegraphics[width=1\textwidth]{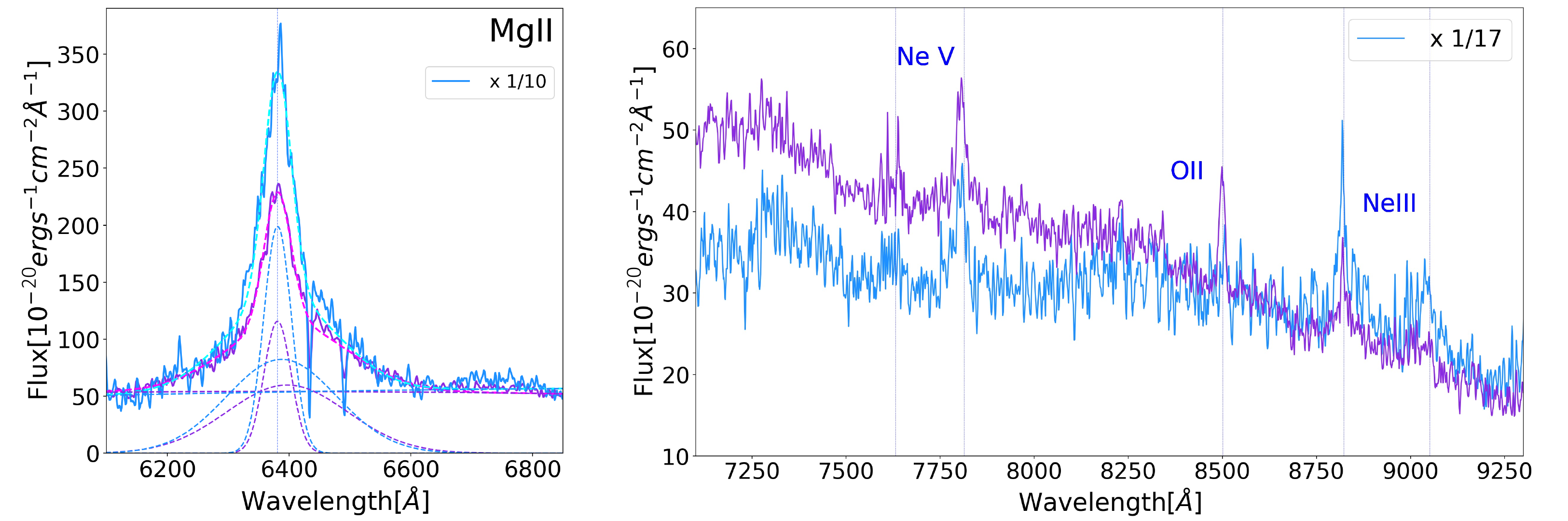}
    \caption{Details of the spectra of J0402-3237, with the same color-coding of Fig \ref{fig:fitLinesJ1059}. Left panel: fit to the MgII line. Right panel: part of the spectrum showing several narrow lines, after multiplying the spectrum of the B component by the factor indicated in the label. Note the clear difference in the observed [OII] emission line and in the shape of the continuum.}
    \label{fig:fitLinesJ0402}
    \end{figure*}
    
\subsubsection{J0402$-$3237}
\label{sec:J0402}
The system in question was previously observed by \cite{Lemon20}, who were not able to provide a reliable classification due to the limited spatial resolution of their data. In fact, our MUSE cubes fully resolve the individual components and allow us to classify the system as a dual AGN.
We de-reddened the spectra of J0402$-$3237 by using the MW attenuation, and noted that the values of E(B$-$V) are markedly different between the two components, namely 0.189 and 0.861. Notably, below 5600~\AA\, the B component (light blue in Fig. \ref{fig:spectra}) is very faint, resulting in a very low signal-to-noise ratio, thus we exclude this portion of the spectrum in the comparison. We tried to improve the match between the two spectra by incorporating a second-degree polynomial to account for potential variations in extinction between the spectra.

Our analysis of the line ratios revealed numerous differences, shown in Fig. \ref{fig:fitLinesJ0402} and quantified in Tab. \ref{tab:features}. Specifically, the EWs of MgII$\lambda2798$, [NeV]$\lambda3426$, [NeIII]$\lambda3869$ and [NeIII]$\lambda3968$ lines are found to be larger in the fainter AGN (i.e., component B, in light blue), while [NeV]$\lambda3346$ and [OII]$\lambda3728$ are more prominent in the brighter one (i.e., the violet one).

Assuming it is a lensed system, we checked for the presence of a lens or the effect of microlensing explained in Sects. \ref{sec:lens_detection} and \ref{sec:microlensing}. \\
J0402$-$3237 is the only case in which the lens is expected to be visible in the white light image. For any possible lens redshift, we conducted simulations of lensing galaxies, yielding magnitudes below G<21.5 (at $z\sim0.7$). All of these simulated lenses are detectable under the specific system configuration of $z_{QSO}=1.28$ and $sep=0.666"$. 
However, despite these efforts, we were unable to detect it, further strengthening our initial interpretation as a true dual AGN.

Using the MmD method of \cite{Sluse07} (priv. comm.) on MgII$\lambda2798$ and [NeV]$\lambda3426$ lines, we find that the microlensing event(s) that could explain the discrepancies between the spectra has a rather strong amplitude, but still not unfeasible.
If the observed chromaticity is interpreted as microlensing, we would require M to be the same for MgII and [NeV] ($M=0.4)$. Anyway, this is not supported by the data because it implies $ \mu=1.65$ ($>1$) at the level of [NeV], but $\mu=0.37$ ($<1$) at the level of MgII. This implies a full microlensing of the [NeV] line, which is not possible. Hence, the observed chromaticity of the spectral ratio as (only) chromatic microlensing is not viable.
The alternative is considering the possibility that the chromaticity in J0402$-$3237 results by differential extinction. In this case the quantity that varies chromatically is M ($ M_{MgII}=0.4$ and $ M_{NeV}=1.4$) while $\mu$ remains constant ($\mu=0.37$).  We find that microlensing of the full [NeV] line would be necessary also in this scenario, but it is unfeasible as narrow lines are not affected by microlensing \citep{Sluse07}. 
Thanks to all the differences described above, the lack of lensing galaxy and the microlensing analysis, we can classify this system as a dual AGN (Table \ref{tab:features}).


\begin{figure}[]
\centering
\includegraphics[width=0.4 \textwidth]
{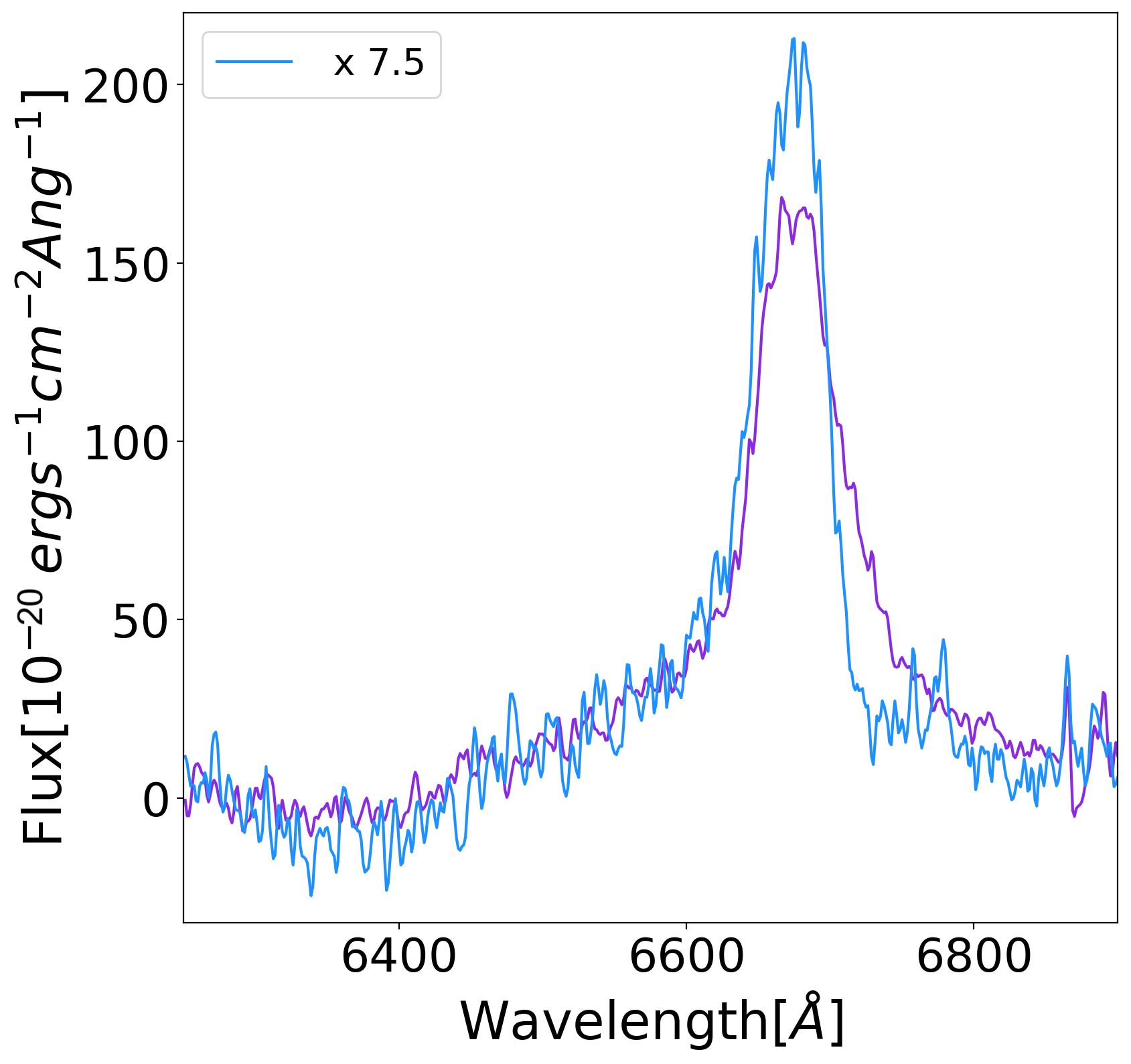}
    \caption{MgII line of J0840$-$0529's components, color-coded as in Fig. \ref{fig:spectra}. To optimize the visualization, the B component has been multiplied by the factors in the label.}
    \label{fig:J0840line}
\end{figure}
\begin{figure}[!h]
\centering
    \includegraphics[width=0.49\textwidth]{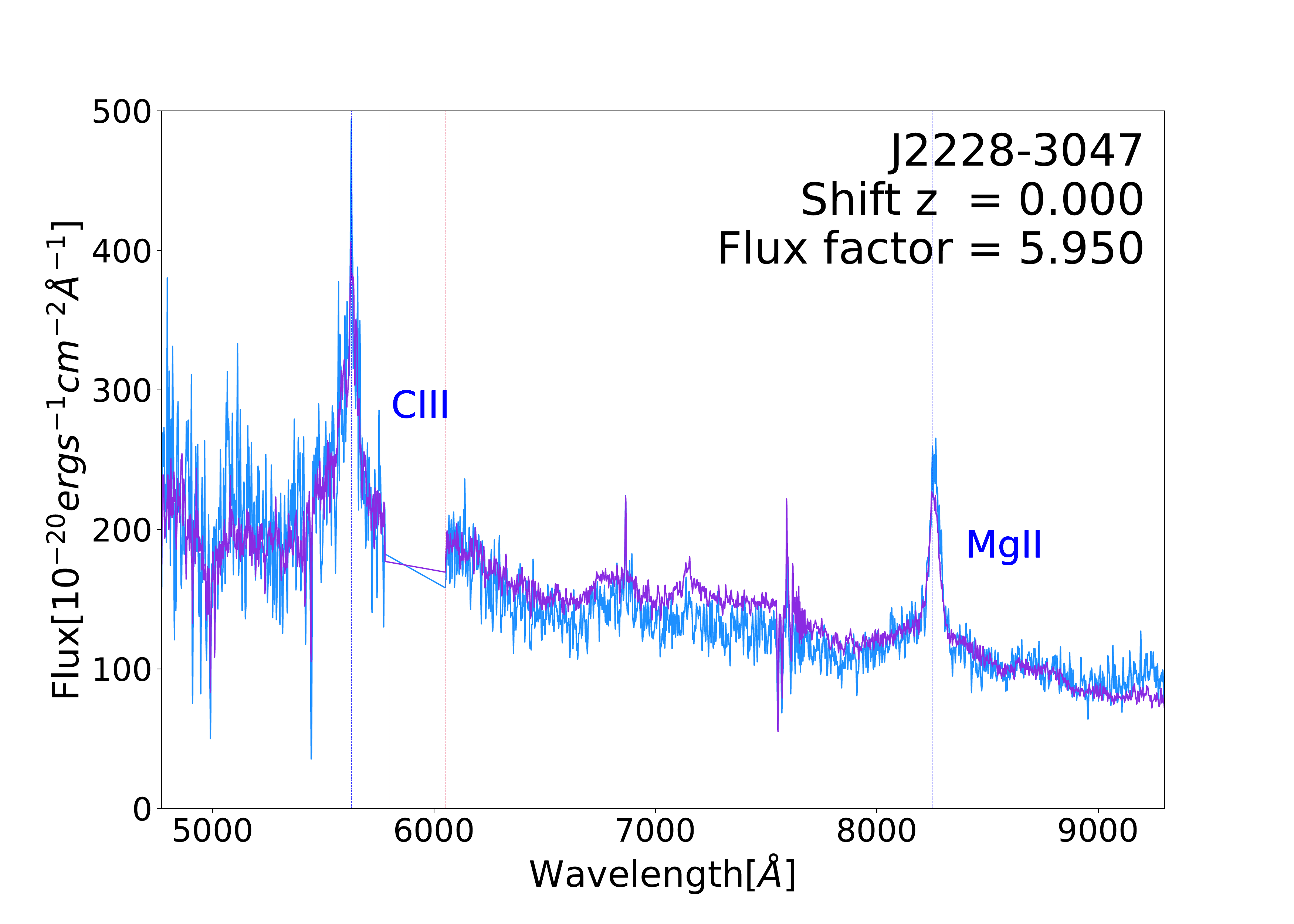}
    \includegraphics[width=0.4\textwidth]{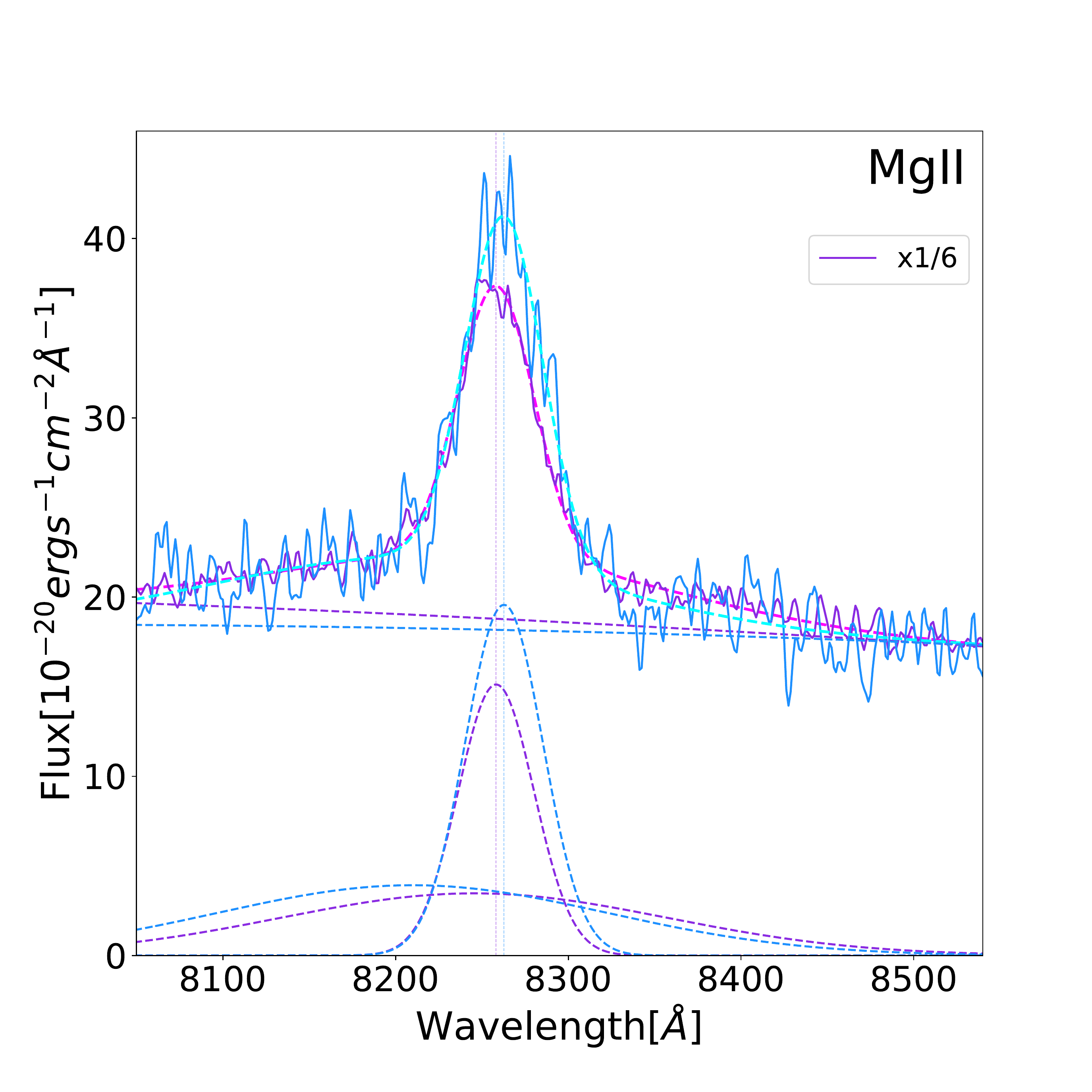}
    \caption{Upper panel: Cross correlation J2228$-$3047\_B (light blue AGN spectra): multiplied in flux and shifted to the same redshift by the quantities indicated in the labels. 
    Lower panel: Fit to the MgII line of both components of J2228-3047. Color-coding as was in Figure \ref{fig:fitLinesJ0402}}
    \label{fig:fitlinesJ2228}
\end{figure}
\subsubsection{J0840$-$0529}
J0840$-$0529 can also be reliably classified as a  dual AGN since, despite having the same redshift, the two spectra exhibit remarkably distinguishable characteristics. For example, the shape of the MgII line is different in the two spectra, as it appears to be narrower in the fainter AGN (component B, light blue one in figure \ref{fig:J0840line}). 

Also the cross-correlation analysis clearly shows significant differences in the CII and MgII emission lines, i.e.; the line flux ratio between MgII and CII is significantly different.
In addition, we calculated the EW of MgII for both the AGN, following the analysis explained in Sec. \ref{sec:fitlines} and fitting the line with two gaussian and one power-law. It turned out to be significantly different,  $24 \pm 3$ for component A vs. $41 \pm 4$ for component B.

As detailed in section \ref{sec:microlensing}, we employed the MmD method of \cite{Sluse07} to examine whether the variations in line shape, EWs, and line ratios could be attributed to microlensing. It is not possible to get a plausible decomposition of MgII. The profile deformations are so pronounced that only the case of microlensing acting stronger on the lines than on the continuum could explain the observed properties. Such an event very unlikely, since microlensing is expected to be stronger on the continuum and less evident on the lines.   
Consequently, the case of a dual scenario with two distinct AGN is more plausible.

\subsubsection{J2228$-$3047} 

Both AGN of J2228$-$3047 display prominent emission lines of MgII and CIII].
Several characteristics suggest that this is a dual system.
We fit the CIII] line with one Gaussian component plus a power law for the continuum,  whereas we added a second Gaussian for the broad-line contribution in the fit of the MgII line (Fig. \ref{fig:fitlinesJ2228}). The fits indicate that the difference in the line central wavelengths in the two AGN is too large to be due to variability of a single source over a few days. We also calculated the EWs of the narrow lines of MgII and CIII] and their line ratio. Values are significantly different between the two components, see Table~\ref{tab:features}.
Furthermore, the cross-correlation analysis also shows intrinsic differences in the EWs of the iron lines. 

Still, this system could be a lensed AGN if microlensing affects the continuum solely through flux magnification and the MgII line through deformation of its shape.
We try to systematically apply the MmD \citep{Sluse07} of J0840--0529 to see if we can get a plausible decomposition of it.
Data are rather noisy, and consequently, the decomposition is uncertain. At the beginning, the data seem to be compatible with a lensing hypothesis implying a microlensing of the continuum $\mu =0.7$ and a small line deformation of MgII (but not of the CIII] complex). This hypothesis necessitates an additional minor differential extinction or intrinsic variability to account for the observed differences (around $20\%$) between CIII] ($M=0.4$) and MgII ($M=0.32$). On the other hand, the absence of a consistent decrease or increase in the flux ratio between the blue and red wings of these lines suggests to us to exclude the lensing hypothesis. This may be harder to explain if $\mu$ is effectively achromatic as was suggested by the 2 lines analyzed. Consequently, we cannot completely dismiss the microlensing hypothesis, particularly given the low S/N.

Nonetheless, the results of the analysis of J2228$-$3047 further suggest the possibility of the duality of this system.

\section{Discussion and conclusions}
\label{sec:summary}

We used spatially resolved, AO-assisted spectroscopy with MUSE-NFM to unveil the nature of 12 multiple AGN candidates selected through the GMP method \citep{Mannucci22}; that is showing multiple peaks in the Gaia light profile, at redshifts between 0.3 and 2.9.
Nine of the targets have been selected from objects with previous spatially unresolved spectroscopy (Sec. \ref{sec:spec_targets}), and 3 from the Gaia QSO candidate catalog (Sec. \ref{sec:phot_targets}).
GMP-selected objects could be either, physically associated dual AGN, or, gravitationally lensed systems, or an AGN projected close to a foreground Galactic star. Spatially resolved observations allow us to
obtain individual spectra for both components of each target 
and derive a reliable classification of the systems.
A novel decomposition routine, based on reproducing the available, spatially unresolved spectrum of the systems with a combination of an AGN and a stellar spectra (see Sect. \ref{sec:deconv}) was used to remove targets with clear hints of the presence of a foreground star.

The spatial resolution of our AO spectra, having PSF FWHM$\sim$0.15", allows us to resolve all the sources.
All the targets in our sample present two components, 
with separations between $0.30\arcsec$ and $0.86\arcsec$ (Fig. \ref{fig:cubes}). This confirms that the GMP method  is very effective in successfully recovering multiple systems with separations below 7 kpc at $z>0.3$.

5 systems turn out to be associations between an AGN and a foreground star. We examine the remaining seven AGN pairs using multiple steps to determine whether the systems are lensed QSOs or dual AGN (Sec. \ref{sec:analysis}), taking into account all the possible sources of differences between the spectra of the different components of a lensed QSO (intrinsic variability, microlensing effects, and differential dust extinctions). For this, we have compared the two spectra by applying different possible distortions due to these effects. We have also compared the position, fluxes, and line profiles of the emission lines, and looked for the possible presence of lens galaxies. The possible effect of microlensing was considered using the MmD decomposition presented in \cite{Sluse07}.

This accurate analysis of the spectra allowed us to unveil the nature of all the seven AGN pairs (Sec. \ref{sec:results}). 
We classified 
4 dual AGN, and 3 lensed QSOs candidate, as was outlined in Table \ref{tab:muse}. Hence, this work significantly increases the number of sub-arcsec separation lensed and dual AGN to date. 

As has been reported by \cite{Mannucci23}, to date there are 15 confirmed dual AGN at z$>$0.5 with projected separations below 7 kpc, including the 4 new sources classified in this work. 
Future observations from the ground (especially with VLT/MUSE, VLT/ERIS, and Keck/OSIRIS) and from space (HST/STIS, JWST) are expected to reveal more of these systems, hence largely increasing the number of confirmed dual AGN, as well as sub-arcsecond lensed systems. 

This sample is still too small to derive meaningful statistical information, but it is rapidly growing.  In the near future, it will be possible to start testing many of the predictions of the models of SMBH merging, such as the SMBH mass function, distribution of separations, fraction of multiple AGN over the general population of quasars, and their evolution with redshift \citep[e.g.,][]{Rosas-Guevara2019, Volonteri21,Chen23b}. 
Moreover, the compact lensed systems that we are discovering will allow us to sample the inner part of the lensing galaxies and measure the mass of the baryonic component in regions where the dark matter component is negligible \citep[e.g.,][]{Smith13}.
\\

\section*{Acknowledgements}
Based on observations collected at the European Southern Observatory under ESO programs [MUSE programs ID: 109.22W5 and 110.23SM; NTT program ID: 109.22W4]

This publication  was produced while attending the PhD program in PhD in Space Science and Technology at the University of Trento, Cycle XXXIX, with the support of a scholarship financed by the Ministerial Decree no. 118 of 2nd march 2023, based on the NRRP - funded by the European Union - NextGenerationEU - Mission 4 "Education and Research", Component 1 "Enhancement of the offer of educational services: from nurseries to universities” - Investment 4.1 “Extension of the number of research doctorates and innovative doctorates for public administration and cultural heritage”.\\
We thank the anonymous referee for his detailed comments to the original manuscript that contributed in improving the text and leading to stronger conclusions.
We are deeply grateful to Dominique Sluse for his assistance in computing the analysis of gravitational microlensing. We thank Alessandro Sonnenfeld for useful discussions.\\
We acknowledge financial contribution from INAF Large Grant “Dual and binary supermassive black holes in the multi-messenger era: from galaxy mergers to gravitational waves” (Bando Ricerca Fondamentale INAF 2022), from the INAF project "VLT-MOONS" CRAM 1.05.03.07, from the French National Research Agency (grant ANR-21-CE31-0026, project MBH\_waves), from the INAF Large Grant 2022 "The metal circle: a new sharp view of the baryon cycle up to Cosmic Dawn with the latest generation IFU facilities".
This work has been partially financed by the European Union with the Next Generation EU plan, through PRIN-MUR project "PROMETEUS" (202223XPZM).
G.V. acknowledges support from the ANID program FONDECYT Postdoctorado 3200802 and the European Union’s HE ERC Starting Grant No. 101040227-WINGS.\\
This work has made use of data from the European Space Agency (ESA) mission {\it Gaia} (\url{https://www.cosmos.esa.int/gaia}), processed by the {\it Gaia}
Data Processing and Analysis Consortium (DPAC,
\url{https://www.cosmos.esa.int/web/gaia/dpac/consortium}). Funding for the DPAC
has been provided by national institutions, in particular the institutions
participating in the {\it Gaia} Multilateral Agreement.

Funding for the Sloan Digital Sky Survey (SDSS) has been provided by the Alfred P. Sloan Foundation, the Participating Institutions, the National Aeronautics and Space Administration, the National Science Foundation, the U.S. Department of Energy, the Japanese Monbukagakusho, and the Max Planck Society. The SDSS Web site is http://www.sdss.org/. The SDSS is managed by the Astrophysical Research Consortium (ARC) for the Participating Institutions. The Participating Institutions are The University of Chicago, Fermilab, the Institute for Advanced Study, the Japan Participation Group, The Johns Hopkins University, Los Alamos National Laboratory, the Max-Planck-Institute for Astronomy (MPIA), the Max-Planck-Institute for Astrophysics (MPA), New Mexico State University, University of Pittsburgh, Princeton University, the United States Naval Observatory, and the University of Washington.

\bibliographystyle{aa}
\bibliography{references} 

\begin{thebibliography}{69}
\expandafter\ifx\csname natexlab\endcsname\relax\def\natexlab#1{#1}\fi

\bibitem[{{Amaro-Seoane} {et~al.}(2023){Amaro-Seoane}, {Andrews}, {Arca Sedda}, {Askar}, {Baghi}, {Balasov}, {Bartos}, {Bavera}, {Bellovary}, {Berry}, {Berti}, {Bianchi}, {Blecha}, {Blondin}, {Bogdanovi{\'c}}, {Boissier}, {Bonetti}, {Bonoli}, {Bortolas}, {Breivik}, {Capelo}, {Caramete}, {Cattorini}, {Charisi}, {Chaty}, {Chen}, {Chru{\'s}li{\'n}ska}, {Chua}, {Church}, {Colpi}, {D'Orazio}, {Danielski}, {Davies}, {Dayal}, {De Rosa}, {Derdzinski}, {Destounis}, {Dotti}, {Du{\r{A}}{\textsterling}an}, {Dvorkin}, {Fabj}, {Foglizzo}, {Ford}, {Fouvry}, {Franchini}, {Fragos}, {Fryer}, {Gaspari}, {Gerosa}, {Graziani}, {Groot}, {Habouzit}, {Haggard}, {Haiman}, {Han}, {Istrate}, {Johansson}, {Khan}, {Kimpson}, {Kokkotas}, {Kong}, {Korol}, {Kremer}, {Kupfer}, {Lamberts}, {Larson}, {Lau}, {Liu}, {Lloyd-Ronning}, {Lodato}, {Lupi}, {Ma}, {Maccarone}, {Mandel}, {Mangiagli}, {Mapelli}, {Mathis}, {Mayer}, {McGee}, {McKernan}, {Miller}, {Mota}, {Mumpower}, {Nasim}, {Nelemans}, {Noble}, {Pacucci}, {Panessa}, {Paschalidis},
  {Pfister}, {Porquet}, {Quenby}, {Ricarte}, {R{\"o}pke}, {Regan}, {Rosswog}, {Ruiter}, {Ruiz}, {Runnoe}, {Schneider}, {Schnittman}, {Secunda}, {Sesana}, {Seto}, {Shao}, {Shapiro}, {Sopuerta}, {Stone}, {Suvorov}, {Tamanini}, {Tamfal}, {Tauris}, {Temmink}, {Tomsick}, {Toonen}, {Torres-Orjuela}, {Toscani}, {Tsokaros}, {Unal}, {V{\'a}zquez-Aceves}, {Valiante}, {van Putten}, {van Roestel}, {Vignali}, {Volonteri}, {Wu}, {Younsi}, {Yu}, {Zane}, {Zwick}, {Antonini}, {Baibhav}, {Barausse}, {Bonilla Rivera}, {Branchesi}, {Branduardi-Raymont}, {Burdge}, {Chakraborty}, {Cuadra}, {Dage}, {Davis}, {de Mink}, {Decarli}, {Doneva}, {Escoffier}, {Gandhi}, {Haardt}, {Lousto}, {Nissanke}, {Nordhaus}, {O'Shaughnessy}, {Portegies Zwart}, {Pound}, {Schussler}, {Sergijenko}, {Spallicci}, {Vernieri}, \& {Vigna-G{\'o}mez}}]{Amaro-Seoane23}
{Amaro-Seoane}, P., {Andrews}, J., {Arca Sedda}, M., {et~al.} 2023, Living Reviews in Relativity, 26, 2

\bibitem[{{Arzoumanian} {et~al.}(2018){Arzoumanian}, {Baker}, {Brazier}, {Burke-Spolaor}, {Chamberlin}, {Chatterjee}, {Christy}, {Cordes}, {Cornish}, {Crawford}, {Thankful Cromartie}, {Crowter}, {DeCesar}, {Demorest}, {Dolch}, {Ellis}, {Ferdman}, {Ferrara}, {Folkner}, {Fonseca}, {Garver-Daniels}, {Gentile}, {Haas}, {Hazboun}, {Huerta}, {Islo}, {Jones}, {Jones}, {Kaplan}, {Kaspi}, {Lam}, {Lazio}, {Levin}, {Lommen}, {Lorimer}, {Luo}, {Lynch}, {Madison}, {McLaughlin}, {McWilliams}, {Mingarelli}, {Ng}, {Nice}, {Park}, {Pennucci}, {Pol}, {Ransom}, {Ray}, {Rasskazov}, {Siemens}, {Simon}, {Spiewak}, {Stairs}, {Stinebring}, {Stovall}, {Swiggum}, {Taylor}, {Vallisneri}, {van Haasteren}, {Vigeland}, {Zhu}, \& {NANOGrav Collaboration}}]{Arzoumanian18}
{Arzoumanian}, Z., {Baker}, P.~T., {Brazier}, A., {et~al.} 2018, \apj, 859, 47

\bibitem[{Bacon {et~al.}(2010)Bacon, Accardo, Adjali, Anwand, Bauer, Biswas, Blaizot, Boudon, Brau-Nogue, Brinchmann, Caillier, Capoani, Carollo, Contini, Couderc, Daguis{\'{e}}, Deiries, Delabre, Dreizler, Dubois, Dupieux, Dupuy, Emsellem, Fechner, Fleischmann, Fran{\c{c}}ois, Gallou, Gharsa, Glindemann, Gojak, Guiderdoni, Hansali, Hahn, Jarno, Kelz, Koehler, Kosmalski, Laurent, Floch, Lilly, Lizon, Loupias, Manescau, Monstein, Nicklas, Olaya, Pares, Pasquini, P{\'{e}}contal-Rousset, Pell{\'{o}}, Petit, Popow, Reiss, Remillieux, Renault, Roth, Rupprecht, Serre, Schaye, Soucail, Steinmetz, Streicher, Stuik, H, Vernet, Weilbacher, Wisotzki, \& Yerle}]{Bacon10}
Bacon, R., Accardo, M., Adjali, L., {et~al.} 2010, in {SPIE} Proceedings, ed. I.~S. McLean, S.~K. Ramsay, \& H.~Takami ({SPIE})

\bibitem[{{Begelman} {et~al.}(1980){Begelman}, {Blandford}, \& {Rees}}]{Begelman80}
{Begelman}, M.~C., {Blandford}, R.~D., \& {Rees}, M.~J. 1980, \nat, 287, 307

\bibitem[{{Bentz} {et~al.}(2013){Bentz}, {Denney}, {Grier}, {Barth}, {Peterson}, {Vestergaard}, {Bennert}, {Canalizo}, {De Rosa}, {Filippenko}, {Gates}, {Greene}, {Li}, {Malkan}, {Pogge}, {Stern}, {Treu}, \& {Woo}}]{Bentz13}
{Bentz}, M.~C., {Denney}, K.~D., {Grier}, C.~J., {et~al.} 2013, \apj, 767, 149

\bibitem[{{Calzetti} {et~al.}(2000){Calzetti}, {Armus}, {Bohlin}, {Kinney}, {Koornneef}, \& {Storchi-Bergmann}}]{Calzetti00}
{Calzetti}, D., {Armus}, L., {Bohlin}, R.~C., {et~al.} 2000, \apj, 533, 682

\bibitem[{{Cardelli} {et~al.}(1989){Cardelli}, {Clayton}, \& {Mathis}}]{Cardelli89}
{Cardelli}, J.~A., {Clayton}, G.~C., \& {Mathis}, J.~S. 1989, \apj, 345, 245

\bibitem[{{Chartas}(2007)}]{Chartas}
{Chartas}, G. 2007, {a Survey of Nal Quasars with High Velocity Outflows}, XMM-Newton Proposal ID 05526702

\bibitem[{{Chen} {et~al.}(2023{\natexlab{a}}){Chen}, {Di Matteo}, {Ni}, {Tremmel}, {DeGraf}, {Shen}, {Holgado}, {Bird}, {Croft}, \& {Feng}}]{Chen23b}
{Chen}, N., {Di Matteo}, T., {Ni}, Y., {et~al.} 2023{\natexlab{a}}, \mnras, 522, 1895

\bibitem[{{Chen}(2021)}]{Chen22b}
{Chen}, Y.-C. 2021, arXiv e-prints, arXiv:2109.06881

\bibitem[{{Chen} {et~al.}(2022){Chen}, {Hwang}, {Shen}, {Liu}, {Zakamska}, {Yang}, \& {Li}}]{Chen22a}
{Chen}, Y.-C., {Hwang}, H.-C., {Shen}, Y., {et~al.} 2022, \apj, 925, 162

\bibitem[{{Chen} {et~al.}(2023{\natexlab{b}}){Chen}, {Liu}, {Foord}, {Shen}, {Oguri}, {Chen}, {Di Matteo}, {Holgado}, {Hwang}, \& {Zakamska}}]{Chen23a}
{Chen}, Y.-C., {Liu}, X., {Foord}, A., {et~al.} 2023{\natexlab{b}}, \nat, 616, 45

\bibitem[{Ciurlo {et~al.}(2023)Ciurlo, Mannucci, Yeh, Amiri, Carniani, Cicone, Cresci, Lusso, Marasco, Marconcini, Marconi, Nardini, Pancino, Rosati, Rubinur, Severgnini, Scialpi, Tozzi, Venturi, Vignali, \& Volonteri}]{Ciurlo23}
Ciurlo, A., Mannucci, F., Yeh, S., {et~al.} 2023, A\&A, 671, L4

\bibitem[{{Colpi}(2014)}]{Colpi14}
{Colpi}, M. 2014, \ssr, 183, 189

\bibitem[{{Croom} {et~al.}(2009){Croom}, {Richards}, {Shanks}, {Boyle}, {Strauss}, {Myers}, {Nichol}, {Pimbblet}, {Ross}, {Schneider}, {Sharp}, \& {Wake}}]{Croom09}
{Croom}, S.~M., {Richards}, G.~T., {Shanks}, T., {et~al.} 2009, \mnras, 399, 1755

\bibitem[{{Delchambre} {et~al.}(2023){Delchambre}, {Bailer-Jones}, {Bellas-Velidis}, {Drimmel}, {Garabato}, {Carballo}, {Hatzidimitriou}, {Marshall}, {Andrae}, {Dafonte}, {Livanou}, {Fouesneau}, {Licata}, {Lindstr{\o}m}, {Manteiga}, {Robin}, {Silvelo}, {Abreu Aramburu}, {{\'A}lvarez}, {Bakker}, {Bijaoui}, {Brouillet}, {Brugaletta}, {Burlacu}, {Casamiquela}, {Chaoul}, {Chiavassa}, {Contursi}, {Cooper}, {Creevey}, {Dapergolas}, {de Laverny}, {Demouchy}, {Dharmawardena}, {Edvardsson}, {Fr{\'e}mat}, {Garc{\'\i}a-Lario}, {Garc{\'\i}a-Torres}, {Gavel}, {Gomez}, {Gonz{\'a}lez-Santamar{\'\i}a}, {Heiter}, {Jean-Antoine Piccolo}, {Kontizas}, {Kordopatis}, {Korn}, {Lanzafame}, {Lebreton}, {Lobel}, {Lorca}, {Magdaleno Romeo}, {Marocco}, {Mary}, {Nicolas}, {Ordenovic}, {Pailler}, {Palicio}, {Pallas-Quintela}, {Panem}, {Pichon}, {Poggio}, {Recio-Blanco}, {Riclet}, {Rybizki}, {Santove{\~n}a}, {Sarro}, {Schultheis}, {Segol}, {Slezak}, {Smart}, {Sordo}, {Soubiran}, {S{\"u}veges}, {Th{\'e}venin}, {Torralba Elipe}, {Ulla},
  {Utrilla}, {Vallenari}, {van Dillen}, {Zhao}, \& {Zorec}}]{Delchambre22}
{Delchambre}, L., {Bailer-Jones}, C.~A.~L., {Bellas-Velidis}, I., {et~al.} 2023, \aap, 674, A31

\bibitem[{{Einstein}(1936)}]{Einstein36}
{Einstein}, A. 1936, Science, 84, 506

\bibitem[{{Fian} {et~al.}(2018){Fian}, {Guerras}, {Mediavilla}, {Jim{\'e}nez-Vicente}, {Mu{\~n}oz}, {Falco}, {Motta}, \& {Hanslmeier}}]{Fian18}
{Fian}, C., {Guerras}, E., {Mediavilla}, E., {et~al.} 2018, \apj, 859, 50

\bibitem[{{Flesch}(2021)}]{Flesch21}
{Flesch}, E.~W. 2021, arXiv e-prints, arXiv:2105.12985

\bibitem[{{Freudling} {et~al.}(2013){Freudling}, {Romaniello}, {Bramich}, {Ballester}, {Forchi}, {Garc{\'\i}a-Dabl{\'o}}, {Moehler}, \& {Neeser}}]{Freudling13}
{Freudling}, W., {Romaniello}, M., {Bramich}, D.~M., {et~al.} 2013, \aap, 559, A96

\bibitem[{{Gaia Collaboration} {et~al.}(2023){Gaia Collaboration}, {Bailer-Jones}, {Teyssier}, {Delchambre}, {Ducourant}, {Garabato}, {Hatzidimitriou}, {Klioner}, {Rimoldini}, {Bellas-Velidis}, {Carballo}, {Carnerero}, {Diener}, {Fouesneau}, {Galluccio}, {Gavras}, {Krone-Martins}, {Raiteri}, {Teixeira}, {Brown}, {Vallenari}, {Prusti}, {de Bruijne}, {Arenou}, {Babusiaux}, {Biermann}, {Creevey}, {Evans}, {Eyer}, {Guerra}, {Hutton}, {Jordi}, {Lammers}, {Lindegren}, {Luri}, {Mignard}, {Panem}, {Pourbaix}, {Randich}, {Sartoretti}, {Soubiran}, {Tanga}, {Walton}, {Bastian}, {Drimmel}, {Jansen}, {Katz}, {Lattanzi}, {van Leeuwen}, {Bakker}, {Cacciari}, {Casta{\~n}eda}, {De Angeli}, {Fabricius}, {Fr{\'e}mat}, {Guerrier}, {Heiter}, {Masana}, {Messineo}, {Mowlavi}, {Nicolas}, {Nienartowicz}, {Pailler}, {Panuzzo}, {Riclet}, {Roux}, {Seabroke}, {Sordo}, {Th{\'e}venin}, {Gracia-Abril}, {Portell}, {Altmann}, {Andrae}, {Audard}, {Benson}, {Berthier}, {Blomme}, {Burgess}, {Busonero}, {Busso}, {C{\'a}novas}, {Carry}, {Cellino},
  {Cheek}, {Clementini}, {Damerdji}, {Davidson}, {de Teodoro}, {Nu{\~n}ez Campos}, {Dell'Oro}, {Esquej}, {Fern{\'a}ndez-Hern{\'a}ndez}, {Fraile}, {Garc{\'\i}a-Lario}, {Gosset}, {Haigron}, {Halbwachs}, {Hambly}, {Harrison}, {Hern{\'a}ndez}, {Hestroffer}, {Hodgkin}, {Holl}, {Jan{\ss}en}, {Jevardat de Fombelle}, {Jordan}, {Lanzafame}, {L{\"o}ffler}, {Marchal}, {Marrese}, {Moitinho}, {Muinonen}, {Osborne}, {Pancino}, {Pauwels}, {Recio-Blanco}, {Reyl{\'e}}, {Riello}, {Roegiers}, {Rybizki}, {Sarro}, {Siopis}, {Smith}, {Sozzetti}, {Utrilla}, {van Leeuwen}, {Abbas}, {{\'A}brah{\'a}m}, {Abreu Aramburu}, {Aerts}, {Aguado}, {Ajaj}, {Aldea-Montero}, {Altavilla}, {{\'A}lvarez}, {Alves}, {Anderson}, {Anglada Varela}, {Antoja}, {Baines}, {Baker}, {Balaguer-N{\'u}{\~n}ez}, {Balbinot}, {Balog}, {Barache}, {Barbato}, {Barros}, {Barstow}, {Bartolom{\'e}}, {Bassilana}, {Bauchet}, {Becciani}, {Bellazzini}, {Berihuete}, {Bernet}, {Bertone}, {Bianchi}, {Binnenfeld}, {Blanco-Cuaresma}, {Boch}, {Bombrun}, {Bossini}, {Bouquillon},
  {Bragaglia}, {Bramante}, {Breedt}, {Bressan}, {Brouillet}, {Brugaletta}, {Bucciarelli}, {Burlacu}, {Butkevich}, {Buzzi}, {Caffau}, {Cancelliere}, {Cantat-Gaudin}, {Carlucci}, {Carrasco}, {Casamiquela}, {Castellani}, {Castro-Ginard}, {Chaoul}, {Charlot}, {Chemin}, {Chiaramida}, {Chiavassa}, {Chornay}, {Comoretto}, {Contursi}, {Cooper}, {Cornez}, {Cowell}, {Crifo}, {Cropper}, {Crosta}, {Crowley}, {Dafonte}, {Dapergolas}, {David}, {de Laverny}, {De Luise}, {De March}, {De Ridder}, {de Souza}, {de Torres}, {del Peloso}, {del Pozo}, {Delbo}, {Delgado}, {Delisle}, {Demouchy}, {Dharmawardena}, {Diakite}, {Distefano}, {Dolding}, {Enke}, {Fabre}, {Fabrizio}, {Faigler}, {Fedorets}, {Fernique}, {Figueras}, {Fournier}, {Fouron}, {Fragkoudi}, {Gai}, {Garcia-Gutierrez}, {Garcia-Reinaldos}, {Garc{\'\i}a-Torres}, {Garofalo}, {Gavel}, {Gerlach}, {Geyer}, {Giacobbe}, {Gilmore}, {Girona}, {Giuffrida}, {Gomel}, {Gomez}, {Gonz{\'a}lez-N{\'u}{\~n}ez}, {Gonz{\'a}lez-Santamar{\'\i}a}, {Gonz{\'a}lez-Vidal}, {Granvik}, {Guillout},
  {Guiraud}, {Guti{\'e}rrez-S{\'a}nchez}, {Guy}, {Hauser}, {Haywood}, {Helmer}, {Helmi}, {Sarmiento}, {Hidalgo}, {Hilger}, {H{\l}adczuk}, {Hobbs}, {Holland}, {Huckle}, {Jardine}, {Jasniewicz}, {Jean-Antoine Piccolo}, {Jim{\'e}nez-Arranz}, {Juaristi Campillo}, {Julbe}, {Karbevska}, {Kervella}, {Khanna}, {Kontizas}, {Kordopatis}, {Korn}, {K{\'o}sp{\'a}l}, {Kostrzewa-Rutkowska}, {Kruszy{\'n}ska}, {Kun}, {Laizeau}, {Lambert}, {Lanza}, {Lasne}, {Le Campion}, {Lebreton}, {Lebzelter}, {Leccia}, {Leclerc}, {Lecoeur-Taibi}, {Liao}, {Licata}, {Lindstr{\o}m}, {Lister}, {Livanou}, {Lobel}, {Lorca}, {Loup}, {Madrero Pardo}, {Magdaleno Romeo}, {Managau}, {Mann}, {Manteiga}, {Marchant}, {Marconi}, {Marcos}, {Marcos Santos}, {Mar{\'\i}n Pina}, {Marinoni}, {Marocco}, {Marshall}, {Martin Polo}, {Mart{\'\i}n-Fleitas}, {Marton}, {Mary}, {Masip}, {Massari}, {Mastrobuono-Battisti}, {Mazeh}, {McMillan}, {Messina}, {Michalik}, {Millar}, {Mints}, {Molina}, {Molinaro}, {Moln{\'a}r}, {Monari}, {Mongui{\'o}}, {Montegriffo}, {Montero},
  {Mor}, {Mora}, {Morbidelli}, {Morel}, {Morris}, {Muraveva}, {Murphy}, {Musella}, {Nagy}, {Noval}, {Oca{\~n}a}, {Ogden}, {Ordenovic}, {Osinde}, {Pagani}, {Pagano}, {Palaversa}, {Palicio}, {Pallas-Quintela}, {Panahi}, {Payne-Wardenaar}, {Pe{\~n}alosa Esteller}, {Penttil{\"a}}, {Pichon}, {Piersimoni}, {Pineau}, {Plachy}, {Plum}, {Poggio}, {Pr{\v{s}}a}, {Pulone}, {Racero}, {Ragaini}, {Rainer}, {Ramos}, {Ramos-Lerate}, {Re Fiorentin}, {Regibo}, {Richards}, {Rios Diaz}, {Ripepi}, {Riva}, {Rix}, {Rixon}, {Robichon}, {Robin}, {Robin}, {Roelens}, {Rogues}, {Rohrbasser}, {Romero-G{\'o}mez}, {Rowell}, {Royer}, {Ruz Mieres}, {Rybicki}, {Sadowski}, {S{\'a}ez N{\'u}{\~n}ez}, {Sagrist{\`a} Sell{\'e}s}, {Sahlmann}, {Salguero}, {Samaras}, {Sanchez Gimenez}, {Sanna}, {Santove{\~n}a}, {Sarasso}, {Schultheis}, {Sciacca}, {Segol}, {Segovia}, {S{\'e}gransan}, {Semeux}, {Shahaf}, {Siddiqui}, {Siebert}, {Siltala}, {Silvelo}, {Slezak}, {Slezak}, {Smart}, {Snaith}, {Solano}, {Solitro}, {Souami}, {Souchay}, {Spagna}, {Spina},
  {Spoto}, {Steele}, {Steidelm{\"u}ller}, {Stephenson}, {S{\"u}veges}, {Surdej}, {Szabados}, {Szegedi-Elek}, {Taris}, {Taylor}, {Tolomei}, {Tonello}, {Torra}, {Torra}, {Torralba Elipe}, {Trabucchi}, {Tsounis}, {Turon}, {Ulla}, {Unger}, {Vaillant}, {van Dillen}, {van Reeven}, {Vanel}, {Vecchiato}, {Viala}, {Vicente}, {Voutsinas}, {Weiler}, {Wevers}, {Wyrzykowski}, {Yoldas}, {Yvard}, {Zhao}, {Zorec}, {Zucker}, \& {Zwitter}}]{Bailer-Jones23}
{Gaia Collaboration}, {Bailer-Jones}, C.~A.~L., {Teyssier}, D., {et~al.} 2023, \aap, 674, A41

\bibitem[{{Gaia Collaboration} {et~al.}(2021){Gaia Collaboration}, {Brown}, {Vallenari}, {Prusti}, {de Bruijne}, {Babusiaux}, {Biermann}, {Creevey}, {Evans}, {Eyer}, {Hutton}, {Jansen}, {Jordi}, {Klioner}, {Lammers}, {Lindegren}, {Luri}, {Mignard}, {Panem}, {Pourbaix}, {Randich}, {Sartoretti}, {Soubiran}, {Walton}, {Arenou}, {Bailer-Jones}, {Bastian}, {Cropper}, {Drimmel}, {Katz}, {Lattanzi}, {van Leeuwen}, {Bakker}, {Cacciari}, {Casta{\~n}eda}, {De Angeli}, {Ducourant}, {Fabricius}, {Fouesneau}, {Fr{\'e}mat}, {Guerra}, {Guerrier}, {Guiraud}, {Jean-Antoine Piccolo}, {Masana}, {Messineo}, {Mowlavi}, {Nicolas}, {Nienartowicz}, {Pailler}, {Panuzzo}, {Riclet}, {Roux}, {Seabroke}, {Sordo}, {Tanga}, {Th{\'e}venin}, {Gracia-Abril}, {Portell}, {Teyssier}, {Altmann}, {Andrae}, {Bellas-Velidis}, {Benson}, {Berthier}, {Blomme}, {Brugaletta}, {Burgess}, {Busso}, {Carry}, {Cellino}, {Cheek}, {Clementini}, {Damerdji}, {Davidson}, {Delchambre}, {Dell'Oro}, {Fern{\'a}ndez-Hern{\'a}ndez}, {Galluccio}, {Garc{\'\i}a-Lario},
  {Garcia-Reinaldos}, {Gonz{\'a}lez-N{\'u}{\~n}ez}, {Gosset}, {Haigron}, {Halbwachs}, {Hambly}, {Harrison}, {Hatzidimitriou}, {Heiter}, {Hern{\'a}ndez}, {Hestroffer}, {Hodgkin}, {Holl}, {Jan{\ss}en}, {Jevardat de Fombelle}, {Jordan}, {Krone-Martins}, {Lanzafame}, {L{\"o}ffler}, {Lorca}, {Manteiga}, {Marchal}, {Marrese}, {Moitinho}, {Mora}, {Muinonen}, {Osborne}, {Pancino}, {Pauwels}, {Petit}, {Recio-Blanco}, {Richards}, {Riello}, {Rimoldini}, {Robin}, {Roegiers}, {Rybizki}, {Sarro}, {Siopis}, {Smith}, {Sozzetti}, {Ulla}, {Utrilla}, {van Leeuwen}, {van Reeven}, {Abbas}, {Abreu Aramburu}, {Accart}, {Aerts}, {Aguado}, {Ajaj}, {Altavilla}, {{\'A}lvarez}, {{\'A}lvarez Cid-Fuentes}, {Alves}, {Anderson}, {Anglada Varela}, {Antoja}, {Audard}, {Baines}, {Baker}, {Balaguer-N{\'u}{\~n}ez}, {Balbinot}, {Balog}, {Barache}, {Barbato}, {Barros}, {Barstow}, {Bartolom{\'e}}, {Bassilana}, {Bauchet}, {Baudesson-Stella}, {Becciani}, {Bellazzini}, {Bernet}, {Bertone}, {Bianchi}, {Blanco-Cuaresma}, {Boch}, {Bombrun}, {Bossini},
  {Bouquillon}, {Bragaglia}, {Bramante}, {Breedt}, {Bressan}, {Brouillet}, {Bucciarelli}, {Burlacu}, {Busonero}, {Butkevich}, {Buzzi}, {Caffau}, {Cancelliere}, {C{\'a}novas}, {Cantat-Gaudin}, {Carballo}, {Carlucci}, {Carnerero}, {Carrasco}, {Casamiquela}, {Castellani}, {Castro-Ginard}, {Castro Sampol}, {Chaoul}, {Charlot}, {Chemin}, {Chiavassa}, {Cioni}, {Comoretto}, {Cooper}, {Cornez}, {Cowell}, {Crifo}, {Crosta}, {Crowley}, {Dafonte}, {Dapergolas}, {David}, {David}, {de Laverny}, {De Luise}, {De March}, {De Ridder}, {de Souza}, {de Teodoro}, {de Torres}, {del Peloso}, {del Pozo}, {Delbo}, {Delgado}, {Delgado}, {Delisle}, {Di Matteo}, {Diakite}, {Diener}, {Distefano}, {Dolding}, {Eappachen}, {Edvardsson}, {Enke}, {Esquej}, {Fabre}, {Fabrizio}, {Faigler}, {Fedorets}, {Fernique}, {Fienga}, {Figueras}, {Fouron}, {Fragkoudi}, {Fraile}, {Franke}, {Gai}, {Garabato}, {Garcia-Gutierrez}, {Garc{\'\i}a-Torres}, {Garofalo}, {Gavras}, {Gerlach}, {Geyer}, {Giacobbe}, {Gilmore}, {Girona}, {Giuffrida}, {Gomel}, {Gomez},
  {Gonzalez-Santamaria}, {Gonz{\'a}lez-Vidal}, {Granvik}, {Guti{\'e}rrez-S{\'a}nchez}, {Guy}, {Hauser}, {Haywood}, {Helmi}, {Hidalgo}, {Hilger}, {H{\l}adczuk}, {Hobbs}, {Holland}, {Huckle}, {Jasniewicz}, {Jonker}, {Juaristi Campillo}, {Julbe}, {Karbevska}, {Kervella}, {Khanna}, {Kochoska}, {Kontizas}, {Kordopatis}, {Korn}, {Kostrzewa-Rutkowska}, {Kruszy{\'n}ska}, {Lambert}, {Lanza}, {Lasne}, {Le Campion}, {Le Fustec}, {Lebreton}, {Lebzelter}, {Leccia}, {Leclerc}, {Lecoeur-Taibi}, {Liao}, {Licata}, {Lindstr{\o}m}, {Lister}, {Livanou}, {Lobel}, {Madrero Pardo}, {Managau}, {Mann}, {Marchant}, {Marconi}, {Marcos Santos}, {Marinoni}, {Marocco}, {Marshall}, {Martin Polo}, {Mart{\'\i}n-Fleitas}, {Masip}, {Massari}, {Mastrobuono-Battisti}, {Mazeh}, {McMillan}, {Messina}, {Michalik}, {Millar}, {Mints}, {Molina}, {Molinaro}, {Moln{\'a}r}, {Montegriffo}, {Mor}, {Morbidelli}, {Morel}, {Morris}, {Mulone}, {Munoz}, {Muraveva}, {Murphy}, {Musella}, {Noval}, {Ord{\'e}novic}, {Orr{\`u}}, {Osinde}, {Pagani}, {Pagano},
  {Palaversa}, {Palicio}, {Panahi}, {Pawlak}, {Pe{\~n}alosa Esteller}, {Penttil{\"a}}, {Piersimoni}, {Pineau}, {Plachy}, {Plum}, {Poggio}, {Poretti}, {Poujoulet}, {Pr{\v{s}}a}, {Pulone}, {Racero}, {Ragaini}, {Rainer}, {Raiteri}, {Rambaux}, {Ramos}, {Ramos-Lerate}, {Re Fiorentin}, {Regibo}, {Reyl{\'e}}, {Ripepi}, {Riva}, {Rixon}, {Robichon}, {Robin}, {Roelens}, {Rohrbasser}, {Romero-G{\'o}mez}, {Rowell}, {Royer}, {Rybicki}, {Sadowski}, {Sagrist{\`a} Sell{\'e}s}, {Sahlmann}, {Salgado}, {Salguero}, {Samaras}, {Sanchez Gimenez}, {Sanna}, {Santove{\~n}a}, {Sarasso}, {Schultheis}, {Sciacca}, {Segol}, {Segovia}, {S{\'e}gransan}, {Semeux}, {Shahaf}, {Siddiqui}, {Siebert}, {Siltala}, {Slezak}, {Smart}, {Solano}, {Solitro}, {Souami}, {Souchay}, {Spagna}, {Spoto}, {Steele}, {Steidelm{\"u}ller}, {Stephenson}, {S{\"u}veges}, {Szabados}, {Szegedi-Elek}, {Taris}, {Tauran}, {Taylor}, {Teixeira}, {Thuillot}, {Tonello}, {Torra}, {Torra}, {Turon}, {Unger}, {Vaillant}, {van Dillen}, {Vanel}, {Vecchiato}, {Viala}, {Vicente},
  {Voutsinas}, {Weiler}, {Wevers}, {Wyrzykowski}, {Yoldas}, {Yvard}, {Zhao}, {Zorec}, {Zucker}, {Zurbach}, \& {Zwitter}}]{Brown21}
{Gaia Collaboration}, {Brown}, A.~G.~A., {Vallenari}, A., {et~al.} 2021, \aap, 649, A1

\bibitem[{{Gaia Collaboration} {et~al.}(2016){Gaia Collaboration}, {Prusti}, {de Bruijne}, {Brown}, {Vallenari}, {Babusiaux}, {Bailer-Jones}, {Bastian}, {Biermann}, {Evans}, {Eyer}, {Jansen}, {Jordi}, {Klioner}, {Lammers}, {Lindegren}, {Luri}, {Mignard}, {Milligan}, {Panem}, {Poinsignon}, {Pourbaix}, {Randich}, {Sarri}, {Sartoretti}, {Siddiqui}, {Soubiran}, {Valette}, {van Leeuwen}, {Walton}, {Aerts}, {Arenou}, {Cropper}, {Drimmel}, {H{\o}g}, {Katz}, {Lattanzi}, {O'Mullane}, {Grebel}, {Holland}, {Huc}, {Passot}, {Bramante}, {Cacciari}, {Casta{\~n}eda}, {Chaoul}, {Cheek}, {De Angeli}, {Fabricius}, {Guerra}, {Hern{\'a}ndez}, {Jean-Antoine-Piccolo}, {Masana}, {Messineo}, {Mowlavi}, {Nienartowicz}, {Ord{\'o}{\~n}ez-Blanco}, {Panuzzo}, {Portell}, {Richards}, {Riello}, {Seabroke}, {Tanga}, {Th{\'e}venin}, {Torra}, {Els}, {Gracia-Abril}, {Comoretto}, {Garcia-Reinaldos}, {Lock}, {Mercier}, {Altmann}, {Andrae}, {Astraatmadja}, {Bellas-Velidis}, {Benson}, {Berthier}, {Blomme}, {Busso}, {Carry}, {Cellino}, {Clementini},
  {Cowell}, {Creevey}, {Cuypers}, {Davidson}, {De Ridder}, {de Torres}, {Delchambre}, {Dell'Oro}, {Ducourant}, {Fr{\'e}mat}, {Garc{\'\i}a-Torres}, {Gosset}, {Halbwachs}, {Hambly}, {Harrison}, {Hauser}, {Hestroffer}, {Hodgkin}, {Huckle}, {Hutton}, {Jasniewicz}, {Jordan}, {Kontizas}, {Korn}, {Lanzafame}, {Manteiga}, {Moitinho}, {Muinonen}, {Osinde}, {Pancino}, {Pauwels}, {Petit}, {Recio-Blanco}, {Robin}, {Sarro}, {Siopis}, {Smith}, {Smith}, {Sozzetti}, {Thuillot}, {van Reeven}, {Viala}, {Abbas}, {Abreu Aramburu}, {Accart}, {Aguado}, {Allan}, {Allasia}, {Altavilla}, {{\'A}lvarez}, {Alves}, {Anderson}, {Andrei}, {Anglada Varela}, {Antiche}, {Antoja}, {Ant{\'o}n}, {Arcay}, {Atzei}, {Ayache}, {Bach}, {Baker}, {Balaguer-N{\'u}{\~n}ez}, {Barache}, {Barata}, {Barbier}, {Barblan}, {Baroni}, {Barrado y Navascu{\'e}s}, {Barros}, {Barstow}, {Becciani}, {Bellazzini}, {Bellei}, {Bello Garc{\'\i}a}, {Belokurov}, {Bendjoya}, {Berihuete}, {Bianchi}, {Bienaym{\'e}}, {Billebaud}, {Blagorodnova}, {Blanco-Cuaresma}, {Boch},
  {Bombrun}, {Borrachero}, {Bouquillon}, {Bourda}, {Bouy}, {Bragaglia}, {Breddels}, {Brouillet}, {Br{\"u}semeister}, {Bucciarelli}, {Budnik}, {Burgess}, {Burgon}, {Burlacu}, {Busonero}, {Buzzi}, {Caffau}, {Cambras}, {Campbell}, {Cancelliere}, {Cantat-Gaudin}, {Carlucci}, {Carrasco}, {Castellani}, {Charlot}, {Charnas}, {Charvet}, {Chassat}, {Chiavassa}, {Clotet}, {Cocozza}, {Collins}, {Collins}, {Costigan}, {Crifo}, {Cross}, {Crosta}, {Crowley}, {Dafonte}, {Damerdji}, {Dapergolas}, {David}, {David}, {De Cat}, {de Felice}, {de Laverny}, {De Luise}, {De March}, {de Martino}, {de Souza}, {Debosscher}, {del Pozo}, {Delbo}, {Delgado}, {Delgado}, {di Marco}, {Di Matteo}, {Diakite}, {Distefano}, {Dolding}, {Dos Anjos}, {Drazinos}, {Dur{\'a}n}, {Dzigan}, {Ecale}, {Edvardsson}, {Enke}, {Erdmann}, {Escolar}, {Espina}, {Evans}, {Eynard Bontemps}, {Fabre}, {Fabrizio}, {Faigler}, {Falc{\~a}o}, {Farr{\`a}s Casas}, {Faye}, {Federici}, {Fedorets}, {Fern{\'a}ndez-Hern{\'a}ndez}, {Fernique}, {Fienga}, {Figueras}, {Filippi},
  {Findeisen}, {Fonti}, {Fouesneau}, {Fraile}, {Fraser}, {Fuchs}, {Furnell}, {Gai}, {Galleti}, {Galluccio}, {Garabato}, {Garc{\'\i}a-Sedano}, {Gar{\'e}}, {Garofalo}, {Garralda}, {Gavras}, {Gerssen}, {Geyer}, {Gilmore}, {Girona}, {Giuffrida}, {Gomes}, {Gonz{\'a}lez-Marcos}, {Gonz{\'a}lez-N{\'u}{\~n}ez}, {Gonz{\'a}lez-Vidal}, {Granvik}, {Guerrier}, {Guillout}, {Guiraud}, {G{\'u}rpide}, {Guti{\'e}rrez-S{\'a}nchez}, {Guy}, {Haigron}, {Hatzidimitriou}, {Haywood}, {Heiter}, {Helmi}, {Hobbs}, {Hofmann}, {Holl}, {Holland}, {Hunt}, {Hypki}, {Icardi}, {Irwin}, {Jevardat de Fombelle}, {Jofr{\'e}}, {Jonker}, {Jorissen}, {Julbe}, {Karampelas}, {Kochoska}, {Kohley}, {Kolenberg}, {Kontizas}, {Koposov}, {Kordopatis}, {Koubsky}, {Kowalczyk}, {Krone-Martins}, {Kudryashova}, {Kull}, {Bachchan}, {Lacoste-Seris}, {Lanza}, {Lavigne}, {Le Poncin-Lafitte}, {Lebreton}, {Lebzelter}, {Leccia}, {Leclerc}, {Lecoeur-Taibi}, {Lemaitre}, {Lenhardt}, {Leroux}, {Liao}, {Licata}, {Lindstr{\o}m}, {Lister}, {Livanou}, {Lobel}, {L{\"o}ffler},
  {L{\'o}pez}, {Lopez-Lozano}, {Lorenz}, {Loureiro}, {MacDonald}, {Magalh{\~a}es Fernandes}, {Managau}, {Mann}, {Mantelet}, {Marchal}, {Marchant}, {Marconi}, {Marie}, {Marinoni}, {Marrese}, {Marschalk{\'o}}, {Marshall}, {Mart{\'\i}n-Fleitas}, {Martino}, {Mary}, {Matijevi{\v{c}}}, {Mazeh}, {McMillan}, {Messina}, {Mestre}, {Michalik}, {Millar}, {Miranda}, {Molina}, {Molinaro}, {Molinaro}, {Moln{\'a}r}, {Moniez}, {Montegriffo}, {Monteiro}, {Mor}, {Mora}, {Morbidelli}, {Morel}, {Morgenthaler}, {Morley}, {Morris}, {Mulone}, {Muraveva}, {Musella}, {Narbonne}, {Nelemans}, {Nicastro}, {Noval}, {Ord{\'e}novic}, {Ordieres-Mer{\'e}}, {Osborne}, {Pagani}, {Pagano}, {Pailler}, {Palacin}, {Palaversa}, {Parsons}, {Paulsen}, {Pecoraro}, {Pedrosa}, {Pentik{\"a}inen}, {Pereira}, {Pichon}, {Piersimoni}, {Pineau}, {Plachy}, {Plum}, {Poujoulet}, {Pr{\v{s}}a}, {Pulone}, {Ragaini}, {Rago}, {Rambaux}, {Ramos-Lerate}, {Ranalli}, {Rauw}, {Read}, {Regibo}, {Renk}, {Reyl{\'e}}, {Ribeiro}, {Rimoldini}, {Ripepi}, {Riva}, {Rixon},
  {Roelens}, {Romero-G{\'o}mez}, {Rowell}, {Royer}, {Rudolph}, {Ruiz-Dern}, {Sadowski}, {Sagrist{\`a} Sell{\'e}s}, {Sahlmann}, {Salgado}, {Salguero}, {Sarasso}, {Savietto}, {Schnorhk}, {Schultheis}, {Sciacca}, {Segol}, {Segovia}, {Segransan}, {Serpell}, {Shih}, {Smareglia}, {Smart}, {Smith}, {Solano}, {Solitro}, {Sordo}, {Soria Nieto}, {Souchay}, {Spagna}, {Spoto}, {Stampa}, {Steele}, {Steidelm{\"u}ller}, {Stephenson}, {Stoev}, {Suess}, {S{\"u}veges}, {Surdej}, {Szabados}, {Szegedi-Elek}, {Tapiador}, {Taris}, {Tauran}, {Taylor}, {Teixeira}, {Terrett}, {Tingley}, {Trager}, {Turon}, {Ulla}, {Utrilla}, {Valentini}, {van Elteren}, {Van Hemelryck}, {van Leeuwen}, {Varadi}, {Vecchiato}, {Veljanoski}, {Via}, {Vicente}, {Vogt}, {Voss}, {Votruba}, {Voutsinas}, {Walmsley}, {Weiler}, {Weingrill}, {Werner}, {Wevers}, {Whitehead}, {Wyrzykowski}, {Yoldas}, {{\v{Z}}erjal}, {Zucker}, {Zurbach}, {Zwitter}, {Alecu}, {Allen}, {Allende Prieto}, {Amorim}, {Anglada-Escud{\'e}}, {Arsenijevic}, {Azaz}, {Balm}, {Beck}, {Bernstein},
  {Bigot}, {Bijaoui}, {Blasco}, {Bonfigli}, {Bono}, {Boudreault}, {Bressan}, {Brown}, {Brunet}, {Bunclark}, {Buonanno}, {Butkevich}, {Carret}, {Carrion}, {Chemin}, {Ch{\'e}reau}, {Corcione}, {Darmigny}, {de Boer}, {de Teodoro}, {de Zeeuw}, {Delle Luche}, {Domingues}, {Dubath}, {Fodor}, {Fr{\'e}zouls}, {Fries}, {Fustes}, {Fyfe}, {Gallardo}, {Gallegos}, {Gardiol}, {Gebran}, {Gomboc}, {G{\'o}mez}, {Grux}, {Gueguen}, {Heyrovsky}, {Hoar}, {Iannicola}, {Isasi Parache}, {Janotto}, {Joliet}, {Jonckheere}, {Keil}, {Kim}, {Klagyivik}, {Klar}, {Knude}, {Kochukhov}, {Kolka}, {Kos}, {Kutka}, {Lainey}, {LeBouquin}, {Liu}, {Loreggia}, {Makarov}, {Marseille}, {Martayan}, {Martinez-Rubi}, {Massart}, {Meynadier}, {Mignot}, {Munari}, {Nguyen}, {Nordlander}, {Ocvirk}, {O'Flaherty}, {Olias Sanz}, {Ortiz}, {Osorio}, {Oszkiewicz}, {Ouzounis}, {Palmer}, {Park}, {Pasquato}, {Peltzer}, {Peralta}, {P{\'e}turaud}, {Pieniluoma}, {Pigozzi}, {Poels}, {Prat}, {Prod'homme}, {Raison}, {Rebordao}, {Risquez}, {Rocca-Volmerange}, {Rosen},
  {Ruiz-Fuertes}, {Russo}, {Sembay}, {Serraller Vizcaino}, {Short}, {Siebert}, {Silva}, {Sinachopoulos}, {Slezak}, {Soffel}, {Sosnowska}, {Strai{\v{z}}ys}, {ter Linden}, {Terrell}, {Theil}, {Tiede}, {Troisi}, {Tsalmantza}, {Tur}, {Vaccari}, {Vachier}, {Valles}, {Van Hamme}, {Veltz}, {Virtanen}, {Wallut}, {Wichmann}, {Wilkinson}, {Ziaeepour}, \& {Zschocke}}]{Prusti16}
{Gaia Collaboration}, {Prusti}, T., {de Bruijne}, J.~H.~J., {et~al.} 2016, \aap, 595, A1

\bibitem[{{Glikman} {et~al.}(2023){Glikman}, {Langgin}, {Johnstone}, {Yoon}, {Comerford}, {Simmons}, {Stacey}, {Lacy}, \& {O'Meara}}]{Glikman23}
{Glikman}, E., {Langgin}, R., {Johnstone}, M.~A., {et~al.} 2023, \apjl, 951, L18

\bibitem[{{Green}(2006)}]{Green06}
{Green}, P.~J. 2006, \apj, 644, 733

\bibitem[{{Grier} {et~al.}(2019){Grier}, {Shen}, {Horne}, {Brandt}, {Trump}, {Hall}, {Kinemuchi}, {Starkey}, {Schneider}, {Ho}, {Homayouni}, {I-Hsiu Li}, {McGreer}, {Peterson}, {Bizyaev}, {Chen}, {Dawson}, {Eftekharzadeh}, {Guo}, {Jia}, {Jiang}, {Kneib}, {Li}, {Li}, {Nie}, {Oravetz}, {Oravetz}, {Pan}, {Petitjean}, {Ponder}, {Rogerson}, {Vivek}, {Zhang}, \& {Zou}}]{Grier19}
{Grier}, C.~J., {Shen}, Y., {Horne}, K., {et~al.} 2019, \apj, 887, 38

\bibitem[{{Gross} {et~al.}(2023){Gross}, {Chen}, {Foord}, {Liu}, {Shen}, {Oguri}, {Goulding}, {Hwang}, {Zakamska}, {Ma}, \& {Nolan}}]{Gross23}
{Gross}, A.~C., {Chen}, Y.-C., {Foord}, A., {et~al.} 2023, \apj, 956, 117

\bibitem[{{Hutsem{\'e}kers} {et~al.}(2010){Hutsem{\'e}kers}, {Borguet}, {Sluse}, {Riaud}, \& {Anguita}}]{Hutsemékers10}
{Hutsem{\'e}kers}, D., {Borguet}, B., {Sluse}, D., {Riaud}, P., \& {Anguita}, T. 2010, \aap, 519, A103

\bibitem[{{Hwang} {et~al.}(2020){Hwang}, {Shen}, {Zakamska}, \& {Liu}}]{Hwang20}
{Hwang}, H.-C., {Shen}, Y., {Zakamska}, N., \& {Liu}, X. 2020, \apj, 888, 73

\bibitem[{{Jim{\'e}nez-Vicente} {et~al.}(2015){Jim{\'e}nez-Vicente}, {Mediavilla}, {Kochanek}, \& {Mu{\~n}oz}}]{jimenezvicente2015}
{Jim{\'e}nez-Vicente}, J., {Mediavilla}, E., {Kochanek}, C.~S., \& {Mu{\~n}oz}, J.~A. 2015, \apj, 806, 251

\bibitem[{{Kelley} {et~al.}(2017){Kelley}, {Blecha}, \& {Hernquist}}]{Kelley17}
{Kelley}, L.~Z., {Blecha}, L., \& {Hernquist}, L. 2017, \mnras, 464, 3131

\bibitem[{{Kochanek} {et~al.}(2006){Kochanek}, {Mochejska}, {Morgan}, \& {Stanek}}]{Kochanek06}
{Kochanek}, C.~S., {Mochejska}, B., {Morgan}, N.~D., \& {Stanek}, K.~Z. 2006, \apjl, 637, L73

\bibitem[{{Krone-Martins} {et~al.}(2019){Krone-Martins}, {Graham}, {Stern}, {Djorgovski}, {Delchambre}, {Ducourant}, {Teixeira}, {Drake}, {Scarano}, {Surdej}, {Galluccio}, {Jalan}, {Wertz}, {Kl{\"u}ter}, {Mignard}, {Spindola-Duarte}, {Dobie}, {Slezak}, {Sluse}, {Murphy}, {Boehm}, {Nierenberg}, {Bastian}, {Wambsganss}, \& {LeCampion}}]{KroneMartins19}
{Krone-Martins}, A., {Graham}, M.~J., {Stern}, D., {et~al.} 2019, arXiv e-prints, arXiv:1912.08977

\bibitem[{{Larkin} {et~al.}(2006){Larkin}, {Barczys}, {Krabbe}, {Adkins}, {Aliado}, {Amico}, {Brims}, {Campbell}, {Canfield}, {Gasaway}, {Honey}, {Iserlohe}, {Johnson}, {Kress}, {LaFreniere}, {Magnone}, {Magnone}, {McElwain}, {Moon}, {Quirrenbach}, {Skulason}, {Song}, {Spencer}, {Weiss}, \& {Wright}}]{Larkin06}
{Larkin}, J., {Barczys}, M., {Krabbe}, A., {et~al.} 2006, \nar, 50, 362

\bibitem[{{Lemon} {et~al.}(2020){Lemon}, {Auger}, {McMahon}, {Anguita}, {Apostolovski}, {Chen}, {Fassnacht}, {Melo}, {Motta}, {Shajib}, {Treu}, {Agnello}, {Buckley-Geer}, {Schechter}, {Birrer}, {Collett}, {Courbin}, {Rusu}, {Abbott}, {Allam}, {Annis}, {Avila}, {Bertin}, {Brooks}, {Burke}, {Carnero Rosell}, {Carrasco Kind}, {Carretero}, {Costanzi}, {da Costa}, {De Vicente}, {Desai}, {Eifler}, {Flaugher}, {Frieman}, {Garc{\'\i}a-Bellido}, {Gaztanaga}, {Gerdes}, {Gruen}, {Gruendl}, {Gschwend}, {Gutierrez}, {Honscheid}, {James}, {Kim}, {Krause}, {Kuehn}, {Kuropatkin}, {Lahav}, {Lima}, {Lin}, {Maia}, {March}, {Marshall}, {Menanteau}, {Miquel}, {Palmese}, {Paz-Chinch{\'o}n}, {Plazas}, {Roodman}, {Sanchez}, {Schubnell}, {Serrano}, {Smith}, {Soares-Santos}, {Suchyta}, {Tarle}, \& {Walker}}]{Lemon20}
{Lemon}, C., {Auger}, M.~W., {McMahon}, R., {et~al.} 2020, \mnras, 494, 3491

\bibitem[{{Lemon} {et~al.}(2024){Lemon}, {Courbin}, {More}, {Schechter}, {Ca{\~n}ameras}, {Delchambre}, {Leung}, {Shu}, {Spiniello}, {Hezaveh}, {Kl{\"u}ter}, \& {McMahon}}]{Lemon23}
{Lemon}, C., {Courbin}, F., {More}, A., {et~al.} 2024, \ssr, 220, 23

\bibitem[{{Lemon} {et~al.}(2019){Lemon}, {Auger}, \& {McMahon}}]{Lemon19}
{Lemon}, C.~A., {Auger}, M.~W., \& {McMahon}, R.~G. 2019, \mnras, 483, 4242

\bibitem[{{Lemon} {et~al.}(2017){Lemon}, {Auger}, {McMahon}, \& {Koposov}}]{Lemon17}
{Lemon}, C.~A., {Auger}, M.~W., {McMahon}, R.~G., \& {Koposov}, S.~E. 2017, \mnras, 472, 5023

\bibitem[{{Lewis} {et~al.}(2013){Lewis}, {Bonaccini Calia}, {Buzzoni}, {Duhoux}, {Fischer}, {Guidolin}, {Hintershuster}, {Holzloehner}, {Jolley}, {Pfrommer}, {Popovic}, {Alvarez}, {Beltran}, {Girard}, \& {Gonte}}]{Lewis13}
{Lewis}, S., {Bonaccini Calia}, D., {Buzzoni}, B., {et~al.} 2013, in Proceedings of the Third AO4ELT Conference, ed. S.~{Esposito} \& L.~{Fini}, 119

\bibitem[{{Lindegren} \& {Dravins}(2021)}]{Lindegren21}
{Lindegren}, L. \& {Dravins}, D. 2021, \aap, 652, A45

\bibitem[{{Longhetti} \& {Saracco}(2009)}]{Longhetti09}
{Longhetti}, M. \& {Saracco}, P. 2009, \mnras, 394, 774

\bibitem[{{Lyke} {et~al.}(2020){Lyke}, {Higley}, {McLane}, {Schurhammer}, {Myers}, {Ross}, {Dawson}, {Chabanier}, {Martini}, {Busca}, {Mas des Bourboux}, {Salvato}, {Streblyanska}, {Zarrouk}, {Burtin}, {Anderson}, {Bautista}, {Bizyaev}, {Brandt}, {Brinkmann}, {Brownstein}, {Comparat}, {Green}, {de la Macorra}, {Mu{\~n}oz Guti{\'e}rrez}, {Hou}, {Newman}, {Palanque-Delabrouille}, {P{\^a}ris}, {Percival}, {Petitjean}, {Rich}, {Rossi}, {Schneider}, {Smith}, {Vivek}, \& {Weaver}}]{Lyke20}
{Lyke}, B.~W., {Higley}, A.~N., {McLane}, J.~N., {et~al.} 2020, \apjs, 250, 8

\bibitem[{{Mannucci} {et~al.}(2022){Mannucci}, {Pancino}, {Belfiore}, {Cicone}, {Ciurlo}, {Cresci}, {Lusso}, {Marasco}, {Marconi}, {Nardini}, {Pinna}, {Severgnini}, {Saracco}, {Tozzi}, \& {Yeh}}]{Mannucci22}
{Mannucci}, F., {Pancino}, E., {Belfiore}, F., {et~al.} 2022, Nature Astronomy, 6, 1185

\bibitem[{{Mannucci} {et~al.}(2023){Mannucci}, {Scialpi}, {Ciurlo}, {Yeh}, {Marconcini}, {Tozzi}, {Cresci}, {Marconi}, {Amiri}, {Belfiore}, {Carniani}, {Cicone}, {Nardini}, {Pancino}, {Rubinur}, {Severgnini}, {Ulivi}, {Venturi}, {Vignali}, {Volonteri}, {Pinna}, {Rossi}, {Puglisi}, {Agapito}, {Plantet}, {Ghose}, {Carbonaro}, {Xompero}, {Grani}, {Esposito}, {Power}, {Guerra Ramon}, {Lefebvre}, {Cavallaro}, {Davies}, {Riccardi}, {Macintosh}, {Taylor}, {Dolci}, {Baruffolo}, {Feuchtgruber}, {Kravchenko}, {Rau}, {Sturm}, {Wiezorrek}, {Dallilar}, \& {Kenworthy}}]{Mannucci23}
{Mannucci}, F., {Scialpi}, M., {Ciurlo}, A., {et~al.} 2023, \aap, 680, A53

\bibitem[{{Massey} {et~al.}(2010){Massey}, {Kitching}, \& {Richard}}]{Massey10}
{Massey}, R., {Kitching}, T., \& {Richard}, J. 2010, Reports on Progress in Physics, 73, 086901

\bibitem[{{Mayer} {et~al.}(2007){Mayer}, {Kazantzidis}, {Madau}, {Colpi}, {Quinn}, \& {Wadsley}}]{Mayer07}
{Mayer}, L., {Kazantzidis}, S., {Madau}, P., {et~al.} 2007, Science, 316, 1874

\bibitem[{{Mosquera} {et~al.}(2013){Mosquera}, {Kochanek}, {Chen}, {Dai}, {Blackburne}, \& {Chartas}}]{mosquera2013}
{Mosquera}, A.~M., {Kochanek}, C.~S., {Chen}, B., {et~al.} 2013, \apj, 769, 53

\bibitem[{{Narayan} \& {Bartelmann}(1996)}]{Narayan96}
{Narayan}, R. \& {Bartelmann}, M. 1996, arXiv e-prints, 9606001

\bibitem[{Perna {et~al.}(2023)Perna, Arribas, Marshall, D'Eugenio, Übler, Bunker, Charlot, Carniani, Jakobsen, Maiolino, Pino, Willott, Böker, Circosta, Cresci, Curti, Husemann, Kumari, Lamperti, Pérez-González, \& Scholtz}]{Perna23}
Perna, M., Arribas, S., Marshall, M., {et~al.} 2023, The ultradense, interacting environment of a dual AGN at z $\sim$ 3.3 revealed by JWST/NIRSpec IFS

\bibitem[{{Prince} {et~al.}(2023){Prince}, {Zaja{\v{c}}ek}, {Panda}, {Hryniewicz}, {Kumar Jaiswal}, {Czerny}, {Trzcionkowski}, {Bronikowski}, {Ra{\l}owski}, {Sobrino Figaredo}, {Martinez-Aldama}, {{\'S}niegowska}, {{\'S}redzi{\'n}ska}, {Bilicki}, {Naddaf}, {Pandey}, {Haas}, {Sarna}, {Pietrzy{\'n}ski}, {Karas}, {Olejak}, {Przy{\l}uski}, {Sefako}, {Genade}, {Worters}, {Koz{\l}owski}, \& {Udalski}}]{Prince23}
{Prince}, R., {Zaja{\v{c}}ek}, M., {Panda}, S., {et~al.} 2023, \aap, 678, A189

\bibitem[{{Rosas-Guevara} {et~al.}(2019){Rosas-Guevara}, {Bower}, {McAlpine}, {Bonoli}, \& {Tissera}}]{Rosas-Guevara2019}
{Rosas-Guevara}, Y.~M., {Bower}, R.~G., {McAlpine}, S., {Bonoli}, S., \& {Tissera}, P.~B. 2019, \mnras, 483, 2712

\bibitem[{{Satyapal} {et~al.}(2014){Satyapal}, {Ellison}, {McAlpine}, {Hickox}, {Patton}, \& {Mendel}}]{Ellison14}
{Satyapal}, S., {Ellison}, S.~L., {McAlpine}, W., {et~al.} 2014, \mnras, 441, 1297

\bibitem[{Sedda {et~al.}(2023)Sedda, Naoz, \& Kocsis}]{Sedda23}
Sedda, M.~A., Naoz, S., \& Kocsis, B. 2023, Universe, 9, 138

\bibitem[{{Shen} {et~al.}(2023){Shen}, {Hwang}, {Oguri}, {Chen}, {Di Matteo}, {Ni}, {Bird}, {Zakamska}, {Liu}, {Chen}, \& {Kratter}}]{Shen23b}
{Shen}, Y., {Hwang}, H.-C., {Oguri}, M., {et~al.} 2023, \apj, 943, 38

\bibitem[{{Shen} {et~al.}(2019){Shen}, {Hwang}, {Zakamska}, \& {Liu}}]{Shen19}
{Shen}, Y., {Hwang}, H.-C., {Zakamska}, N., \& {Liu}, X. 2019, \apjl, 885, L4

\bibitem[{{Skrutskie} {et~al.}(2006){Skrutskie}, {Cutri}, {Stiening}, {Weinberg}, {Schneider}, {Carpenter}, {Beichman}, {Capps}, {Chester}, {Elias}, {Huchra}, {Liebert}, {Lonsdale}, {Monet}, {Price}, {Seitzer}, {Jarrett}, {Kirkpatrick}, {Gizis}, {Howard}, {Evans}, {Fowler}, {Fullmer}, {Hurt}, {Light}, {Kopan}, {Marsh}, {McCallon}, {Tam}, {Van Dyk}, \& {Wheelock}}]{skrutskie06}
{Skrutskie}, M.~F., {Cutri}, R.~M., {Stiening}, R., {et~al.} 2006, \aj, 131, 1163

\bibitem[{{Sluse} {et~al.}(2007){Sluse}, {Claeskens}, {Hutsemekers}, \& {Surdej}}]{Sluse07}
{Sluse}, D., {Claeskens}, J.~F., {Hutsemekers}, D., \& {Surdej}, J. 2007, \aap, 468, 885

\bibitem[{{Sluse} {et~al.}(2012){Sluse}, {Hutsem{\'e}kers}, {Courbin}, {Meylan}, \& {Wambsganss}}]{Sluse12}
{Sluse}, D., {Hutsem{\'e}kers}, D., {Courbin}, F., {Meylan}, G., \& {Wambsganss}, J. 2012, \aap, 544, A62

\bibitem[{{Smith} \& {Lucey}(2013)}]{Smith13}
{Smith}, R.~J. \& {Lucey}, J.~R. 2013, \mnras, 434, 1964

\bibitem[{{Str{\"o}bele} {et~al.}(2012){Str{\"o}bele}, {La Penna}, {Arsenault}, {Conzelmann}, {Delabre}, {Duchateau}, {Dorn}, {Fedrigo}, {Hubin}, {Quentin}, {Jolley}, {Kiekebusch}, {Kirchbauer}, {Klein}, {Kolb}, {Kuntschner}, {Le Louarn}, {Lizon}, {Madec}, {Pettazzi}, {Soenke}, {Tordo}, {Vernet}, \& {Muradore}}]{Strobele12}
{Str{\"o}bele}, S., {La Penna}, P., {Arsenault}, R., {et~al.} 2012, in Society of Photo-Optical Instrumentation Engineers (SPIE) Conference Series, Vol. 8447, Adaptive Optics Systems III, ed. B.~L. {Ellerbroek}, E.~{Marchetti}, \& J.-P. {V{\'e}ran}, 844737

\bibitem[{{Temple} {et~al.}(2021){Temple}, {Hewett}, \& {Banerji}}]{Temple21}
{Temple}, M.~J., {Hewett}, P.~C., \& {Banerji}, M. 2021, \mnras, 508, 737

\bibitem[{{van der Wel} {et~al.}(2014){van der Wel}, {Franx}, {van Dokkum}, {Skelton}, {Momcheva}, {Whitaker}, {Brammer}, {Bell}, {Rix}, {Wuyts}, {Ferguson}, {Holden}, {Barro}, {Koekemoer}, {Chang}, {McGrath}, {H{\"a}ussler}, {Dekel}, {Behroozi}, {Fumagalli}, {Leja}, {Lundgren}, {Maseda}, {Nelson}, {Wake}, {Patel}, {Labb{\'e}}, {Faber}, {Grogin}, \& {Kocevski}}]{vanderWel14}
{van der Wel}, A., {Franx}, M., {van Dokkum}, P.~G., {et~al.} 2014, \apj, 788, 28

\bibitem[{{Vanden Berk} {et~al.}(2001){Vanden Berk}, {Richards}, {Bauer}, {Strauss}, {Schneider}, {Heckman}, {York}, {Hall}, {Fan}, {Knapp}, {Anderson}, {Annis}, {Bahcall}, {Bernardi}, {Briggs}, {Brinkmann}, {Brunner}, {Burles}, {Carey}, {Castander}, {Connolly}, {Crocker}, {Csabai}, {Doi}, {Finkbeiner}, {Friedman}, {Frieman}, {Fukugita}, {Gunn}, {Hennessy}, {Ivezi{\'c}}, {Kent}, {Kunszt}, {Lamb}, {Leger}, {Long}, {Loveday}, {Lupton}, {Meiksin}, {Merelli}, {Munn}, {Newberg}, {Newcomb}, {Nichol}, {Owen}, {Pier}, {Pope}, {Rockosi}, {Schlegel}, {Siegmund}, {Smee}, {Snir}, {Stoughton}, {Stubbs}, {SubbaRao}, {Szalay}, {Szokoly}, {Tremonti}, {Uomoto}, {Waddell}, {Yanny}, \& {Zheng}}]{Vandenberk01}
{Vanden Berk}, D.~E., {Richards}, G.~T., {Bauer}, A., {et~al.} 2001, \aj, 122, 549

\bibitem[{{Volonteri} {et~al.}(2022){Volonteri}, {Pfister}, {Beckmann}, {Dotti}, {Dubois}, {Massonneau}, {Musoke}, \& {Tremmel}}]{Volonteri21}
{Volonteri}, M., {Pfister}, H., {Beckmann}, R., {et~al.} 2022, \mnras, 514, 640

\bibitem[{{Wambsganss}(1998)}]{Wambsganss98}
{Wambsganss}, J. 1998, Living Reviews in Relativity, 1, 12

\bibitem[{{Wambsganss}(2006)}]{Wambsganss06}
{Wambsganss}, J. 2006, arXiv e-prints, 0604278

\bibitem[{{Weilbacher} {et~al.}(2020){Weilbacher}, {Palsa}, {Streicher}, {Bacon}, {Urrutia}, {Wisotzki}, {Conseil}, {Husemann}, {Jarno}, {Kelz}, {P{\'e}contal-Rousset}, {Richard}, {Roth}, {Selman}, \& {Vernet}}]{Weilbacher20}
{Weilbacher}, P.~M., {Palsa}, R., {Streicher}, O., {et~al.} 2020, \aap, 641, A28

\bibitem[{{Wong} {et~al.}(2020){Wong}, {Suyu}, {Chen}, {Rusu}, {Millon}, {Sluse}, {Bonvin}, {Fassnacht}, {Taubenberger}, {Auger}, {Birrer}, {Chan}, {Courbin}, {Hilbert}, {Tihhonova}, {Treu}, {Agnello}, {Ding}, {Jee}, {Komatsu}, {Shajib}, {Sonnenfeld}, {Blandford}, {Koopmans}, {Marshall}, \& {Meylan}}]{Wong20}
{Wong}, K.~C., {Suyu}, S.~H., {Chen}, G. C.~F., {et~al.} 2020, \mnras, 498, 1420

\bibitem[{{Yan} {et~al.}(2019){Yan}, {Chen}, {Lazarz}, {Bizyaev}, {Maraston}, {Stringfellow}, {McCarthy}, {Meneses-Goytia}, {Law}, {Thomas}, {Falcon Barroso}, {S{\'a}nchez-Gallego}, {Schlafly}, {Zheng}, {Argudo-Fern{\'a}ndez}, {Beaton}, {Beers}, {Bershady}, {Blanton}, {Brownstein}, {Bundy}, {Chambers}, {Cherinka}, {De Lee}, {Drory}, {Galbany}, {Holtzman}, {Imig}, {Kaiser}, {Kinemuchi}, {Liu}, {Luo}, {Magnier}, {Majewski}, {Nair}, {Oravetz}, {Oravetz}, {Pan}, {Sobeck}, {Stassun}, {Talbot}, {Tremonti}, {Waters}, {Weijmans}, {Wilhelm}, {Zasowski}, {Zhao}, \& {Zhao}}]{Yan19}
{Yan}, R., {Chen}, Y., {Lazarz}, D., {et~al.} 2019, \apj, 883, 175

\end{thebibliography}


\begin{appendix}
\section{Estimation of QSO near-IR colors from Gaia photometry }
The TT star wavefront sensor used by MUSE-NFM can only work on-axis, i.e, using the science target itself as TT star. For this reason it is necessary to select targets that are bright enough at near-IR wavelengths to obtian a good AO correction. In most cases, the QSO are too faint to have measured J- and H-band magnitudes by 2MASS  \citep{skrutskie06} or any other survey. For this reason, we have derived an estimate of these magnitudes from a linear combination of the observed Rp and Bp magnitudes from Gaia. We have considered all the QSO with observed J- and H-band magnitudes, mostly from 2MASS, and computed the combination of the two Gaia magnitudes minimizing the difference between the observed and predicted magnitudes. The derived relations, together with their scatter, are:
\begin{equation}
J = Rp - 0.82*(Bp-Rp) - 0.26,  ~~~~~~\sigma=0.27
\label{eq:J}
\end{equation}
\vspace{-0.5cm}
\begin{equation}
H = Rp  - 0.89*(Bp-Rp) -0.92,   ~~~~~~\sigma=0.31
\label{eq:H}
\end{equation}

\begin{figure}[h!]
\centering
\includegraphics[width=0.5\textwidth]{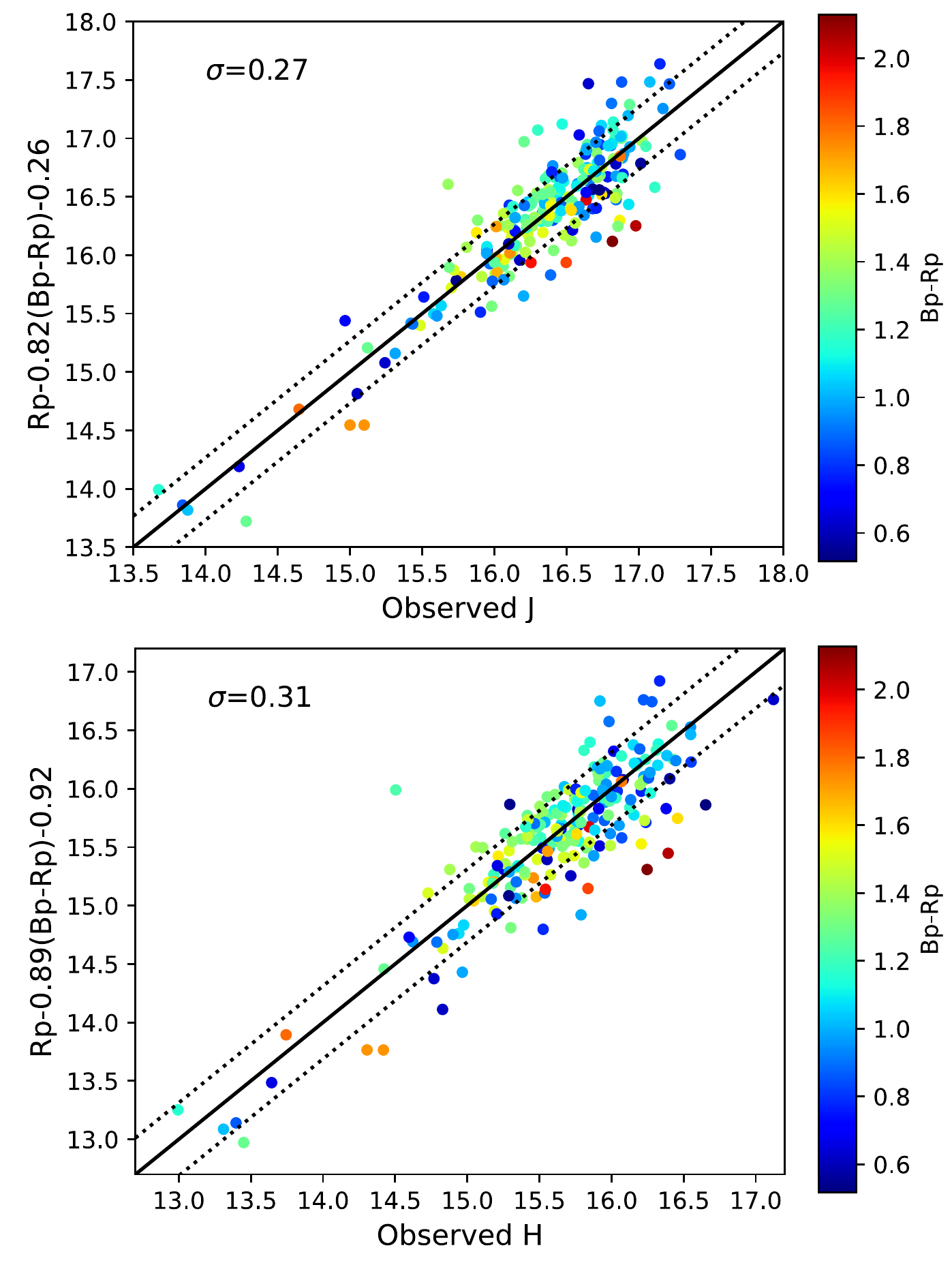}
\caption{Comparison between the observed J- (left panel) and  H- (right panel) magnitudes and the values derived from the Bp- and Rp- Gaia magnitudes using equations \ref{eq:J} and \ref{eq:H}. Each point is color-coded with the observed (Bp-Rp) color. The solid line shows the one-to-one relation, the dotted lines report the observed scatter around this relation.
}
\label{fig:JHcolors}
\end{figure}

These relations are shown in Fig.~\ref{fig:JHcolors}. The observed scatter is small enough to obtain a reliable value of the expected near-IR mag to 
\end{appendix}

\label{lastpage}
\end{document}